\documentclass[11pt,a4paper]{article}
\pdfoutput=1
\usepackage{graphicx}

\usepackage[british]{babel}
\usepackage[utf8]{inputenc}
\usepackage{authblk}
\usepackage{fancyhdr}
\usepackage{amsmath}
\usepackage{amssymb}
\usepackage{yfonts}
\usepackage[colorinlistoftodos]{todonotes}
\usepackage{color}
\usepackage{mathrsfs}
\usepackage{subcaption}
\usepackage{braket}
\usepackage{xfrac}
\usepackage{dsfont}
\usepackage{tikz}
\usetikzlibrary{decorations.pathreplacing,angles,quotes}
\tikzstyle{every picture}+=[remember picture]

\usepackage{breqn}
\usepackage[toc,page]{appendix}
\usepackage{multicol}
\usepackage{hyperref}
\usepackage{bbm}
\hypersetup{
    colorlinks,
    citecolor=black,
    filecolor=black,
    linkcolor=black,
    urlcolor=black
}
\addto\captionsenglish{\renewcommand{\contentsname}%
    {Contents}%
}
\addto\captionsenglish{\renewcommand{\chaptername}%
    {Chapter}%
}
\addto\captionsenglish{\renewcommand{\bibname}%
    {References and Bibliography}%
}
\addto\captionsenglish{}

\newcommand{\CC}{\mathbbm{C}}

\newcommand{\NN}{\mathbbm{N}}

\newcommand{\ZZ}{\mathbbm{Z}}

\def\ll{\left\lgroup}
\def\rr{\right\rgroup}

\def\leq{\leqslant}
\def\geq{\geqslant}

\def\pp{p^{\, \prime}}

\def\ll{ \left\lgroup}
\def\rr{\right\rgroup}

\newcommand{\cB}{\mathcal{B}}

\newcommand{\cF}{\mathcal{F}}

\newcommand{\cH}{\mathcal{H}}

\newcommand{\cM}{\mathcal{M}}
\newcommand{\cN}{\mathcal{N}}
\newcommand{\cO}{\mathcal{O}}

\newcommand{\cV}{\mathcal{V}}
\newcommand{\cW}{\mathcal{W}}

\newcommand{\eps}{\epsilon}

\newcommand{\1}{{\bf 1.}}
\newcommand{\2}{{\bf 2.}}
\newcommand{\3}{{\bf 3.}}

\newcommand{\bea}{\begin{eqnarray}\displaystyle}
\newcommand{\eea}{\end{eqnarray}}

\usepackage[scale=0.82]{geometry}

\begin{document} 
	
\title{Notes on the solutions of Zamolodchikov-type recursion relations 
in Virasoro minimal models}

\author{
Nina Javerzat      \!$^{{\scriptstyle {\, 1            }}}$, 
Raoul Santachiara  \!$^{{\scriptstyle {\, 1            }}}$ and
Omar Foda          \!$^{{\scriptstyle {\, 2            }}}$
          }

\affil{
$^{{\scriptstyle 1}}$ 
LPTMS, CNRS, Univ. Paris-Sud, Université Paris-Saclay, 91405 Orsay, France, 
\newline 
$^{{\scriptstyle 2}}$
School of Mathematics and Statistics, 
University of Melbourne, Parkville, Victoria 3010, Australia
\newline}

\maketitle

\begin{abstract}
We study Virasoro minimal-model 
4-point conformal blocks on the sphere and 
0-point conformal blocks on the torus (the Virasoro characters), 
as solutions of Zamolodchikov-type recursion relations. 
In particular, we study the singularities due to resonances of the dimensions of conformal fields in minimal-model representations, 
that appear in the intermediate steps of solving the recursion relations, but cancel in the 
final results.   
\end{abstract}

\section{Introduction}
\label{section.01}

\subsubsection{The conformal bootstrap and Virasoro conformal blocks} 
In spite of the enormous progress in our understanding of 2D conformal field theories over 
the past 35 years, important classes of two-dimen\-sional conformal field theories remain 
to be discovered, or at least better-understood. An outstanding example is critical percolation, 
which is known to be a CFT with Virasoro central charge $c = 0$, but the correlation functions 
of this CFT remain to be computed. 
Recently, new numerical 2D conformal bootstraps that fully exploit the full 2D local conformal
symmetries were developed and new 2D CFT's were discovered 
\cite{ribault.santachiara, picco.ribault.santachiara}. 
These 2D bootstraps are based on Virasoro conformal blocks, 
and the corresponding 
solutions are functions of infinite-dimensional representations of local  conformal transformations, 
as opposed to bootstraps that are valid in {\it any} dimension, which  
are based on {\it global} conformal blocks, and the corresponding 
solutions are functions of finite-dimensional   representations of global conformal transformations.  

\subsubsection{Zamolodchikov's recursion relation} 
To implement the new 2D bootstraps numerically, one needs to compute the 4-point conformal 
blocks on the sphere efficiently. The most efficient known method to compute 4-point conformal 
blocks on the sphere is Zamolodchikov's recursion relation. In fact, solving the 2D bootstrap 
efficiently is what motivated Al. Zamolodchikov to develop the recursion relations 
in the first place \cite{zamolodchikov.01, zamolodchikov.02}. There are two versions of 
Zamolodchikov's recursion relation, a hypergeometric version \cite{zamolodchikov.01}, and 
an elliptic version \cite{zamolodchikov.02}.  
The elliptic version is particularly efficient, and will be the focus of the present work, 
and to fully understand this recursion relation, we will find it useful to consider a related 
recursion relation for the 1-point conformal block on the torus, and its 0-point conformal 
block limit, which is a Virasoro character.

\subsubsection{The singularities}
The 4-point conformal block on the sphere is a function of six parameters: 
the Virasoro central charge, the conformal dimension of the Virasoro representation that 
flows in the internal channel, and the conformal dimensions of the four external fields. 
The solution of the elliptic recursion relation displays a rich structure of poles. These 
poles are physical in the sense that they correspond to the propagation of  states 
for suitable choices of the central charge and conformal dimensions.
In a numerical 2D bootstrap based on Virasoro conformal blocks that are computed 
using the elliptic recursion relation, one must deal with these poles when exploring the 
space of possible crossing-symmetric CFT solutions. This happens, for example, in studies 
of percolation in the 2D Ising model \cite{picco.ribault.santachiara.02}.
When the central charge is such that one deals with minimal-model conformal blocks, 
additional poles appear. These additional poles are non-physical and appear due to resonances 
of conformal dimensions (see equation (\ref{resonance})) at rational values of the central charge.
This complication requires a careful study of the pole structure of the elliptic recursion 
relation in the case of Virasoro minimal models, which is the aim of the present work.

\subsection{The present work.} 
We study the cancellation of the non-physical 
poles in computations of minimal-model conformal blocks using Zamolodchikov's elliptic 
recursion relation for the 4-point conformal block on the sphere. 
But the 4-point conformal block on the sphere is not the only or the simplest conformal 
block that can be computed using a recursion relation. 
In 2009, Poghossian \cite{poghossian.01}, and independently Fateev and Litvinov
\cite{fateev.litvinov} proposed recursion relations to compute Liouville 1-point conformal 
blocks on the torus. These recursion relations are equivalent \cite{hadasz.jaskolski.suchanek}, 
and we use the Fateev-Litvinov version to 
study minimal-model 1-point functions on the torus, and their 0-point limits 
(when the vertex operator insertion is the identity) which are Virasoro minimal-model 
characters, as the simplest examples of solutions of a Zamolodchikov-type elliptic 
recursion relation. 

\subsection{Outline of contents}
In section \ref{section.02}, 
we recall basic facts related to the Virasoro algebra, its representations, and conformal blocks. 
In section \ref{section.03}, 
we consider the 4-point conformal blocks on the sphere as solutions of the recursion relation, 
study their singularities and their behaviour in the context of the Virasoro generalized 
minimal models and minimal models. 
In section \ref{section.04}, 
we consider the 1-point conformal block on the torus as solutions of the Fateev-Litvinov 
recursion relation.
In section \ref{section.05}, 
we study the solutions of the Fateev-Litvinov recursion 
relations for the Virasoro minimal-model 1-point functions on the torus in the special case 
where the inserted vertex operator is the identity and the 1-point function reduces to the 
character of the irreducible highest-weight representation that flows in the torus.
In appendix \ref{appendix.a}, we include the details of an explicit computation, and 
in appendix \ref{appendix.b}, we include technical details related to coefficients that 
appear in the recursion relations. 

\section{Virasoro algebra, representations and conformal blocks} 
\label{section.02} 
\noindent \textit{We recall basic definitions related to the Virasoro algebra, representation 
theory, and conformal blocks. We refer the reader to the review \cite{Ribault:2014hia}.}

\subsection{The Virasoro algebra, generators and central charge } 
A Virasoro CFT is based on the Virasoro algebra, 
\begin{equation}
[L_{n},L_{m}] = (n - m) L_{n + m} + \frac{c}{12} n (n^2 - 1) \delta_{n + m, 0},
\quad
n, m \in \ZZ,  
\label{virasoro}
\end{equation}

\noindent where the Virasoro generators $L_n, n \in \ZZ$ are the modes of the stress-energy tensor 
and $c$ is the central charge. The Liouville parametrization of the central charge is, 

\begin{equation}
c = 1 + 6 \, Q^{\, 2}, 
\quad 
Q = b + b^{-1}, 
\quad
c \in \CC 
\label{central.charge}
\end{equation}

\subsection{Verma modules} 
Given a highest-weight state $|\Delta \rangle$, with highest weight $\Delta$, 
$L_0 | \Delta \rangle = \Delta \, | \Delta \rangle$, the descendant states 
$L_{- n_1} \cdots L_{- n_N}  |\Delta \rangle$, $n_1 \geq n_2 \geq \cdots \geq n_N$, 
form a basis of the Verma module $\mathcal{V}_{\Delta}$. 
A general element in this basis is $L_{-Y} \, | \, \Delta \, \rangle$, 
labeled by a Young diagram $Y = \ll n_1, \cdots, n_N \rr$, that has $N$ non-zero parts,
and, 

\begin{equation}
L_0 \, | L_{-Y} \Delta \rangle = \ll \Delta+|Y| \rr |L_{-Y} \Delta \rangle, 
\end{equation}

\noindent where $| Y | = \sum_{i=1}^N n_i$ is the number of cells in the Young diagram $Y$. 
Using the state-field correspondence, we use $\Phi_{\Delta} \, (x)$ for the primary field
of conformal dimension $\Delta$, and $L_{-Y} \, \Phi_{\Delta}(x)$ for the descendant fields. 
We parametrize the conformal dimension $\Delta$ by the parameter $Q$,
(\ref{central.charge}), and the charge $\alpha$,

\begin{equation}
\label{DQ}
\Delta = \alpha \ll Q-\alpha\rr
\end{equation}

\subsection{Degenerate representations}

A degenerate representation has a highest weight $\Delta_{m,n}$,

\begin{equation}
\Delta_{m, n} = \alpha_{m, n} \ll Q - \alpha_{m, n} \rr, 
\quad 
\alpha_{\, m, n} = 
- \frac12 \ll m - 1 \rr \, b
- \frac12 \ll n - 1 \rr \, b^{-1}, 
\label{Deltamn}
\end{equation}

\noindent and has a null state  
$| \, \chi_{mn} \, \rangle$ at level $m n$, 
$\langle \, \chi_{\, m n} \, | \, \chi_{\, m n} \rangle = 0$. 
When a representation with highest weight $\Delta_{\, m n}$ appears in the spectrum of 
a given CFT model, two situations can occur.

\subsubsection{Null states vanish}  The corresponding representation module $\mathcal{V}_{m,n}$ 
is the quotient of a reducible Verma module by a non-trivial submodule, 

\begin{equation}
\mathcal{V}_{\, m n} = 
\frac{
\mathcal{V}_{\Delta_{\, m n}}
}{
\mathcal{V}_{\Delta_{\, m n} + m n}
}
\label{rvv}
\end{equation} 

\noindent The representations $\mathcal{V}_{\, m n}$ form the spectrum of the Virasoro 
generalized minimal as well as the minimal models. The vanishing of the null state implies 
the fusion rules. The fusions of products of $\mathcal{V}_{\, m n}$ have simple expressions 
in the parametrization (\ref{DQ}). For instance, the fusion of $\mathcal{V}_{\, m n}$  with a 
Verma module $\mathcal{V}_{\alpha}$ is a sum of $m n$ Verma modules and takes the form,

\begin{equation}
\mathcal{V}_{\, m n} \times \mathcal{V}_\alpha = 
\sum_{i = \frac12 \, \ll 1-m \rr}^{\frac12 \, \ll m - 1 \rr} 
\sum_{j = \frac12 \, \ll 1-n \rr}^{\frac12 \, \ll n - 1 \rr} 
\mathcal{V}_{\alpha + i b + j b^{-1}},
\label{rvsv}
\end{equation}

\noindent where the sums are in steps of $1$.

\subsubsection{Null states do not vanish} Representations with non-vanishing null-states appear 
in Liouville field theory at $c\leq 1$ \cite{ribault.santachiara}, and in computations of 
probabilities of non-local critical objects such as the left-right passage probability 
of an SLE interface. We will not deal with this case in the present work.

\subsection{The Virasoro minimal models}
$\cM_{\, p \, \pp}$, are labeled by two positive co-prime integers $p$ and $\pp$, such that 
$0 < p < \pp$. The space of chiral states of $\cM_{\, p \, \pp}$ is generated by a Virasoro 
algebra with central charge,

\begin{equation}
b = \ll - \, \frac{\pp}{p} \rr^{\, \frac12}
\end{equation}

\noindent The space of chiral states splits into (typically finitely-many) fully-degenerate 
irreducible highest-weight modules $\cV_{\, m n}$ labeled by two integers $m$ and $n$, such 
that $0 < m < p$, and $0 < n < \pp$. From (\ref{Deltamn}), $\Delta_{\, m \, n}$ satisfies 
the negation relation, and the periodicity relation, 

\begin{equation}
\Delta_{\, m \, n} = 
\Delta_{\,    -m,  \,   - n},  
\quad 
\Delta_{\, m, n} = 
\Delta_{\, m + p, \, n + \pp}, 
\label{relation.01.02}
\end{equation}

\noindent which combine to give,

\begin{equation}
\Delta_{\, m, n} =
\Delta_{\, m^{\, \prime}, \, n^{\, \prime}}, 
\quad 
m^{\, \prime} =   p - m, 
\quad 
n^{\, \prime} = \pp - n, 
\label{relation.03}
\end{equation}

\noindent as well as an infinite chain of relations that involve \textit{\lq resonant\rq\,} 
conformal dimensions, 

\begin{equation}
\Delta_{\, m, n} = \Delta_{\, p + m, \, \pp + n} = \cdots
\label{resonance}
\end{equation}

\noindent Two pairs of indexes $\ll m, n \rr$ and $\ll r, s \rr$ are resonant if there 
are linked by a finite chain of transformations (\ref{relation.01.02}).
In this case we use the notation,

\begin{equation}
\ll m, n \rr \leftrightarrow \ll r, s \rr^{\pm}_{l}, 
\end{equation}

\noindent to indicate that there exists an integer $l$ such that,

\begin{equation}
\label{resnotation}
\ll m, n \rr \leftrightarrow \ll r, s \rr^{\pm}_{l} 
\implies    
r = l\,   p \pm  m,\quad 
s = l\, \pp \pm  n,\quad 
l \in \mathbb{N}
\end{equation}

\subsubsection{Remark} In our notation, $0 < p < \pp$, and $b = \sqrt{- \pp / p}$ is pure 
imaginary such that $| b | > 0$. One can think of $| b |$ as the magnitude of the positive 
screening charge $\alpha_{\, +} > 0$. We normally take the negative screening charge 
$\alpha_{\, -} < 0$, and the background charge,  

\begin{equation}
- 2 \alpha_0 = - \, \ll \alpha_+ + \alpha_- \rr, 
\end{equation} 

\noindent that is, the background charge 
can be screened by the sum of a single $\alpha_+$ and a single $\alpha_-$.

\subsection{The Virasoro conformal blocks}
The conformal blocks are special functions of the Virasoro representations. We consider 
the 4-point conformal blocks on the sphere, the 1-point conformal blocks on the torus, 
and the 0-point conformal blocks on the torus, which are Virasoro characters. In all 
generality, these conformal blocks are defined in terms of the $p_{| Y |} \times p_{| Y |}$ 
matrix $S_{| Y |} \ll Y, Y' \rr$ of inner products of descendants at level $| Y |$, where 
$p_{| Y |}$ is the number of partitions of $| Y |$, 

\begin{equation}
S_{| Y |} \ll Y, Y' \rr = 
\langle  L_{-Y} \Delta| L_{-Y'} \Delta \rangle, \quad | Y | = | Y' |, 
\end{equation}

\noindent and the matrix elements,

\begin{equation}
\label{3pt}
\langle L_{-Y_1}\Delta_{1}  | L_{-Y_2}\Phi_{\Delta_{2}}(1) | L_{-Y_3} \Delta_{3} \rangle
\end{equation}

\noindent In Virasoro CFT's, the Shapolavov matrix and the 3-point 
functions are completely determined by the Virasoro algebra (\ref{virasoro}). Note that this is not true anymore for more general conformal chiral algebras such as the $\mathcal{W}^N$ algebras \cite{Belavin:2016wlo}, \cite{Belavin:2016qaa}.

\section{The 4-point conformal blocks on the sphere}
\label{section.03}
\noindent \textit{We outline Zamolodchikov's computation 
of the 4-point conformal block on the sphere, and study its poles.}

\subsection{The 4-point conformal block on the sphere} Global conformal symmetry determines 
the dependences of four-point blocks on three of the four positions $z_i$, and we assume 
$(z_i) = (x, 0, \infty, 1)$. 
The conformal block is a function of six parameters, the central charge, the cross-ratio $x$, 
the conformal dimensions of the four external fields $\Delta_{i}$, $i = 1, \cdots, 4,$ and 
the conformal dimension of the representation that flows in the internal channel 
$\Delta_{\, int}$. In terms of the vertex-operators charges, 

\begin{equation}
\Delta_{i}=\alpha_{i} \ll Q - \alpha_{i} \rr, \quad 
\Delta_{\, int} = \alpha_{\, int} \ll Q - \alpha_{\, int} \rr
\end{equation}

\noindent The 4-point conformal block on the sphere has an $x$-series expansion,  

\begin{equation}
\cB   \ll \mathbf{\Delta}, x \rr = 
x^{ - \Delta_1 - \Delta_2 + \Delta_{\,int}} 
\ll 1 + 
\cB_1 \ll \mathbf{\Delta} \rr \, x +
\cB_2 \ll \mathbf{\Delta} \rr \, x^2 + \cdots
\rr, 
\end{equation}

\noindent where $\mathbf{\Delta}$ for the set of external and internal conformal dimensions 
$\ll \Delta_{1},\Delta_{2},\Delta_{3},\Delta_{4}, \Delta_{\, int} \rr$, and,

\begin{equation}
\label{cbexp}
\cB_{| Y |} \ll \mathbf{\Delta} \rr =
\sum_{\substack{Y, Y^{\, \prime} \\ |Y| = |Y'| }}
S_{| Y |}^{(-1)} \ll Y, Y^{\, \prime} \rr
\langle              
\Delta_2  \, | \Phi_{\, 1} (1) | L_{-Y'} \Delta_{\,int} \, 
\rangle
\langle 
\, L_{-Y} \, \Delta_{\, int} \, | \Phi_{\, 3} (1) | \Delta_4 \, 
\rangle, 
\end{equation}

\noindent where $Y'$ is \textit{any} Young diagram such that $| Y' | = | Y |$, and 
$\Phi_i$ is a primary field of conformal dimension $\Delta_i$.

\subsection{The elliptic recursion relation}
In \cite{zamolodchikov.02}, Zamolodchikov introduced an elliptic recursion relation 
of the same 4-point conformal blocks on the sphere
\footnote{\,
In \cite{poghossian.02}, Poghossian extended Zamolodchikov's elliptic recursion relation 
to a class of $\cW_3$ Toda 4-point conformal blocks on the sphere.
}.
The recursion parameter $q$ is a function of $x$

\begin{equation}
\label{defq}
q = \frac{1}{16} 
\ll x + \frac12 \, x^2+ \frac{21}{64} \, x^3+
\frac{31}{128} \, x^4+ \cO \ll x^5 \rr \rr, 
\end{equation}

\noindent which follows from inverting,

\begin{equation}
x = 
\frac{
\theta^{\, 4}_{\, 2} \ll q \rr
}{
\theta^{\, 4}_{\, 3} \ll q \rr
}, 
\end{equation}

\noindent where $\theta_2 \ll q \rr$ and $\theta_3 \ll q \rr$ are Jacobi theta functions,

\begin{equation}
\label{defell}
\theta_{\, 2} \ll q \rr = \sum_{n = - \infty}^{ \infty} q^{\, \ll n + 1/2\rr^{\, 2}}, \quad
\theta_{\, 3} \ll q \rr = \sum_{n = - \infty}^{ \infty} q^{\, n^{\, 2}},
\end{equation}

\noindent The conformal blocks can be written as,

\begin{multline}
\cB \ll \mathbf{\Delta}, c,x \rr = 
\\
x^{\frac{c - 1}{24} - \Delta_1 - \Delta_2} 
\ll 1-x \rr^{\, \frac{c - 1}{24} -\Delta_2-\Delta_3}
\ll 16 q \rr^{\Delta_{\, int} - \frac{Q^2}{4}}
\theta_3 \ll q \rr^{3 Q^2 - 4 \ll \Delta_1 + \Delta_2 + \Delta_3 + \Delta_4 \rr} 
\times 
 \\ 
\mathcal{H}^{\, \textit{sph}} 
\ll \mathbf{\Delta}_{ext},\Delta_{\, int}, c, x \rr,  
\label{block_ell}
\end{multline}

\noindent where the elliptic variable $q$ and function $\theta_3(q)$ are 
defined in (\ref{defq}) and in (\ref{defell}) and we use 
$\mathbf{\Delta}_{ext}$ for the set of external dimensions 
$\ll \Delta_1, \Delta_2,\Delta_3, \Delta_4 \rr$. 
The analytic structure of the function 
$\cH^{\, \textit{sph}} \ll \mathbf{\Delta}_{ext},\Delta_{\, int}, c, x \rr$ 
is manifest in the following expansion \footnote{Note that some closed form expression has been found in \cite{Perlmutter:2015iya}},

\begin{multline}
\cH^{\, \textit{sph}} \ll \mathbf{\Delta}_{ext},\Delta_{\, int}, c, x \rr = 
1 + 
\sum_{\, rs\geq 1} \ll 16q \rr^{rs}
\frac{R^{\, \textit{sph}}_{\, r, s}\ll \mathbf{\Delta}_{ext},c \rr }{
\Delta_{\, int} - \Delta_{\, r, s}
}\cH^{\, \textit{sph}} \ll \mathbf{\Delta}_{ext},\Delta_{\, r,-s}, c, x \rr, 
\label{H}
\end{multline}

\noindent where,

\begin{equation}
R^{\, \textit{sph}}_{\, m, n} \ll \mathbf{\Delta},c \rr 
= 
\frac{1}{r_{m,n}} \,
P_{\, m, n} \ll \Delta_1,\Delta_2 \rr P_{\, m, n} \ll \Delta_3,\Delta_4  \rr
\label{Rmnsp}
\end{equation}

\noindent The factors $P_{\, m, n}$ carries all dependence in 
$R^{\, \textit{sph}}_{\, m, n} \ll \mathbf{\Delta} \rr$ on the external conformal dimensions 
$\Delta_i, i = 1, \cdots, 4$. It is convenient to parametrize the conformal dimensions in terms 
of the momenta $\lambda_{i}$ and $\lambda_{m,n}$, 

\begin{equation}
\Delta_i = \frac{c-1}{24} + \lambda_{i}^2, \quad \Delta_{m,n}= \frac{c-1}{24} + \lambda_{m,n}^2
\end{equation}

\noindent In terms of these variables, one has,

\begin{multline}
\label{pmn}
P_{\, m, n} \ll \Delta_1,\Delta_2 \rr 
= \prod_{\rho,\sigma} \; 
\ll \lambda_{1}+\lambda_2-\lambda_{\rho,\sigma}\rr
\ll \lambda_{1}-\lambda_2-\lambda_{\rho,\sigma}\rr
\\
\rho=1-m,3-m, \cdots, m-1, 
\quad 
\sigma=1-n,3-n, \cdots, n-1 
\end{multline}
\noindent The factor $r_{m, n}$ is given by,

\begin{multline}
r_{m,n} = - \frac12 \prod_{\rho,\sigma} 2\lambda_{\rho, \sigma},
\\ 
\rho=1-m,2-m, \cdots, m, 
\quad 
\sigma=1-n,2-n, \cdots, n, \quad (\rho,\sigma)\neq (0,0), \;(m,n) 
\label{rmn}
\end{multline}

\subsection{The generalized minimal model} 
When the central charge is non-rational, but a degenerate representation
$\mathcal{V}_{\Delta_{m, n}}$ flows in the channel, the recursion relation 
(\ref{H}) has a pole related to the presence of a null-state at level $m n$ 
in $\mathcal{V}_{\Delta_{m, n}}$, and the corresponding Shapovalov matrix has 
a vanishing eigenvalue that produces the singularity in the expansion (\ref{cbexp}). 

\subsubsection{The regularization $\epsilon$}
We introduce a regularization parameter $\eps$, 

\begin{equation}
\Delta_{\, int}=\Delta^{(\eps)}_{m,n}=\Delta_{m n}+ \eps
\label{Deltamnreg}
\end{equation}

\noindent The limit $\eps \to 0$ in (\ref{H}) exists only if the polynomial 
$P_{m,n}(\Delta_1,\Delta_2)$ and/or  $P_{m,n}(\Delta_3,\Delta_4)$, defined 
in (\ref{Rmnsp}) and in (\ref{pmn}), vanish. Recall that $P_{m,n}(\Delta_1,\Delta_2)$ 
vanishes when ($\cV_{\Delta_1}, \cV_{\Delta_2}, \cV_{m,n})$ satisfy the fusion rules 
(\ref{rvsv}), that is to say when $\alpha_2 =\alpha_{1}+ i b +j b^{-1}$, with 
$i\in \{(1-m)/2,(3-m)/2,\cdots,(m-1)/2\}$ and $j\in \{(1-n)/2,(3-m)/2,\cdots,(n-1)/2\}$.
The generalized minimal model has a non-rational central charge, $c \notin \mathbb{Q}$, 
and a spectrum formed by all the degenerate representations $\mathcal{V}_{m,n}$ with 
$(m,n) \in \mathbb{N}_{+}$. All the fields in the spectrum satisfy the fusion rules 
(\ref{rvsv}), imposed by the condition $\chi_{mn}=0$. The conformal blocks of the 
generalized minimal model can be obtained by using the recursion relation with 
a simple limiting procedure. This  consists in setting,

\begin{equation}
\label{Deltaextreg}
\Delta_i=\Delta^{(\eps_i)}_{r_i,s_i} = \Delta_{r_i,s_i}+\eps_i, \quad i = 1, \cdots, 4, 
\end{equation}

\noindent with $\eps_i\to 0$, $i=1,\cdots,4$ of the same order of $\eps$, 
$\eps_i = \cO  \ll \eps \rr$, and take the limit $\eps\to 0$. Using,

\begin{equation}
\frac{
H^{\, \textit{sph}}_{m, n} \ll \mathbf{\Delta}, c, x \rr 
}{
\Delta_{\, int} - \Delta_{\, m, n}
}\propto \frac{P_{m,n}\ll \Delta^{(\eps)}_{r_1,s_1}\Delta^{(\eps_2)}_{r_2,s_2}\rr P_{m,n}\ll \Delta^{(\eps_3)}_{r_3,s_3}\Delta^{(\eps_4)}_{r_4,s_4}\rr}{\Delta^{\eps}_{m,n}-\Delta_{m,n}}
\sim O\ll \eps \rr,
\end{equation}

\noindent it is straightforward to see that the term 
$\cH^{sph}_{m,n}\ll \mathbf{\Delta}, c, x \rr $in (\ref{H}) do not contribute. In the generalized 
minimal models therefore, the conformal block with $\Delta_{\, int}=\Delta_{m,n}$ is obtained using 
the sum in (\ref{H}) where the term $(r,s)=(m,n)$ is omitted. 
We stress that, in this procedure limit, the final result is independent of the exact relation 
between the regularization parameters $\eps_i$ and $\eps$. The only thing that matters is the fact 
that the $\eps_i$ and $\eps$ are of the same order. As we will see later, this will not be the case 
for the computation of the characters.

\subsection{Minimal-model conformal blocks.}
We address here the problem of how to obtain the conformal blocks of minimal models $\cM_{p,\pp}$ 
from  the recursion relation (\ref{H}). The main observation is that, with respect to the generalized 
minimal models, there are new poles appearing in (\ref{H}). The location of these extra poles do not 
depend neither on the internal channel field nor on the external fields. They originate from the resonances 
in the conformal dimensions that occur when $c\in \mathbb{Q}$. Let us consider for instance one conformal 
block of the Ising minimal model, $\cM_{4,3}$. At level $7$, two terms appear,

\begin{multline}
\frac{R^{\, \textit{sph}}_{2,1}\ll \mathbf{\Delta}_{ext},c \rr }{\Delta_{\, int}-\Delta_{2,1}}\frac{R^{\, \textit{sph}}_{1,5}\ll \mathbf{\Delta}_{ext},c \rr}{\Delta_{2,-1}-\Delta_{1,5}}, 
\quad \text{and}  
\quad 
\frac{R^{\, \textit{sph}}_{1,3}\ll \mathbf{\Delta}_{ext},c \rr }{\Delta_{\, int}-\Delta_{1,3}}\frac{R^{\, \textit{sph}}_{4,1}\ll \mathbf{\Delta}_{ext},c \rr}{\Delta_{-1,3}-\Delta_{4,1}},
\label{Isingex}
\end{multline}

\noindent that are singular due to the fact that at $c=1/2$, $\ll 2,-1\rr \rightarrow\ll 1,5\rr^-$ and 
$\ll -1,3\rr \rightarrow\ll 4,1\rr^-$. Differently from the poles that originate when 
$\Delta_{\, int}=\Delta_{m,n}$, which are related to the null-state at level $mn$, these other 
singularities can be considered an artifact of the recursion relation in the sense that they are not 
related to any special properties of descendant states. In the Appendix we better explain this point 
with an explicit example. 

\subsubsection{The regularization $\epsilon^{\, \prime}$}
In addition to  (\ref{Deltamnreg}), we introduce a regularization parameter 
$\eps^{\, \prime}$ to the central charge $c$, 

\begin{equation}
\label{creg}
b^2=-\frac{\pp}{p} +\eps',
\end{equation}

\noindent that is of the same order of $\eps$, $\eps'=O\ll \eps \rr$.

\subsection{Conjecture} 
We conjecture that, by setting (\ref{Deltamnreg}),(\ref{Deltaextreg}) and (\ref{creg}), the limit 
$\eps \to 0$ in the recursion relation (\ref{H}) exists and provides the correct minimal model 
conformal block. 

\subsection{Further example} 
We have checked that the two terms (\ref{Isingex}) 
combine to  give  a finite contribution. Another example, at level $20$, is the combination of the 
following five singular terms,

\begin{multline}
\label{ising20}
\frac{ R^{\, \textit{sph}}_{1,1} \ll\mathbf{\Delta}_{ext},c \rr}{ \ll\Delta^{(\eps)}_{1,1}  - \Delta_{1, 1} \rr}
\frac{ R^{\, \textit{sph}}_{4,3} \ll \mathbf{\Delta}_{ext},c \rr}{ \ll \Delta_{1, -1} - \Delta_{4, 3} \rr}
\frac{ R^{\, \textit{sph}}_{7,1} \ll \mathbf{\Delta}_{ext},c \rr}{ \ll \Delta_{4, -3} - \Delta_{7, 1} \rr} 
\, + 
\\
\frac{R^{\, \textit{sph}}_{1,1} \ll \mathbf{\Delta}_{ext},c \rr}{\ll \Delta^{(\eps)}_{1,1}   - \Delta_{1, 1} \rr}
\frac{R^{\, \textit{sph}}_{2,5} \ll \mathbf{\Delta}_{ext},c \rr}{\ll \Delta_{1, -1} - \Delta_{2, 5} \rr}
\frac{R^{\, \textit{sph}}_{1,9} \ll \mathbf{\Delta}_{ext},c \rr}{\ll \Delta_{2, -5} - \Delta_{1, 9} \rr} 
\, + 
\\ 
\frac{R^{\, \textit{sph}}_{2,3} \ll \mathbf{\Delta}_{ext},c \rr}{\ll \Delta^{(\eps)}_{1,1}   - \Delta_{2, 3} \rr}
\frac{R^{\, \textit{sph}}_{5,1} \ll \mathbf{\Delta}_{ext},c \rr}{\ll \Delta_{2, -3} - \Delta_{5, 1} \rr}
\frac{R^{\, \textit{sph}}_{1,9} \ll \mathbf{\Delta}_{ext},c \rr}{\ll \Delta_{5, -1} - \Delta_{1, 9} \rr}
\, + 
\\
\frac{R^{\, \textit{sph}}_{2, 3} \ll \mathbf{\Delta}_{ext},c \rr}{\ll \Delta^{(\eps)}_{1,1}   - \Delta_{2, 3} \rr} 
\frac{R^{\, \textit{sph}}_{1, 7} \ll \mathbf{\Delta}_{ext},c \rr}{\ll \Delta_{2, -3} - \Delta_{1, 7} \rr}
\frac{R^{\, \textit{sph}}_{7, 1} \ll \mathbf{\Delta}_{ext},c \rr}{\ll \Delta_{1, -7} - \Delta_{7, 1} \rr}
\, + \,  
\frac{R^{\, \textit{sph}}_{4, 5} \ll \mathbf{\Delta}_{ext},c \rr}{\ll \Delta^{(\eps)}_{1,1}  -  \Delta_{4, 5} \rr}
\end{multline}

\noindent to a finite contribution. If we can predict the singular terms that, at a given level, 
provide finite contributions, we have not been able to obtain a compact formula for these. As 
we will see in the following, we can control the contribution of these type of singularities 
in the computation of a simpler symmetry function, the character.

\section{1-point conformal blocks on the torus}
\label{section.04}
\noindent \textit{We recall the Fateev-Litvinov recursion relation for the 1-point conformal 
block on the torus, and introduce the structure of its poles.}

\subsection{The 1-point conformal block on the torus} The Virasoro 1-point conformal 
block on the torus consists of a single vertex-operator insertion in a torus geometry, 
and a Virasoro irreducible highest-weight representation flows in the single internal 
channel of the torus. This is a function of four parameters, the central charge, the 
torus parameter $q$, the conformal dimension of the external field $\Delta_{\, ext}$, 
and the conformal dimension of the internal channel $\Delta_{\, int}$.
The conformal dimension of the external vertex-operator is,

\begin{equation}
\Delta_{\, ext} = \alpha_{\, ext} \ll Q - \alpha_{\, ext} \rr,
\end{equation}

\noindent and similarly, the conformal dimension of the representation that flows in 
the torus is,

\begin{equation}
\Delta_{\, int} = \alpha_{\, int} \ll Q - \alpha_{\, int} \rr
\end{equation}

\noindent The torus 1-point conformal block has the $q$-series expansion, 

\begin{equation}
\cF   \ll \mathbf{\Delta}, q \rr = 1 + 
\cF_1 \ll \mathbf{\Delta} \rr \, q +
\cF_2 \ll \mathbf{\Delta} \rr \, q^2 + \cdots, 
\end{equation}

\noindent where $\mathbf{\Delta}$ is a pair of conformal dimensions 
$\ll \Delta_{\, ext}, \, \Delta_{\, int} \rr$, and, 

\begin{equation}
\cF_{| Y |} \ll \mathbf{\Delta} \rr = 
\sum_{\substack{Y, Y' \\ |Y| = |Y'| }} S_{| Y |}^{(-1)} \ll Y, Y' \rr 
\langle 
L_{-Y} \Delta_{\, int} 
\, | \,
\Phi_{\, ext} (1) 
\, | \, 
L_{-Y'} \Delta_{\, int} 
\rangle
\end{equation}

\subsection{The recursion relation of Fateev and Litvinov \cite{fateev.litvinov}}
The 1-point conformal block on the torus is,

\begin{equation}
\cF \ll \mathbf{\Delta} \, | \, q \rr =
\frac{q^{1/24}}{\eta(q)} \cH \ll \mathbf{\Delta} \, | \, q \rr, 
\label{onept}
\end{equation}

\noindent where,

\begin{equation}
\frac{q^{\, 1/24}}{\eta (q)} = 
\prod_{i=1}^{\, \infty} \frac{1}{1 - q^{\, i}} =
 1 + q + \cdots + p_{\, N} \, q^{\, N} + \cdots,
 \label{dchar}
\end{equation}

\noindent $p_{\, N}$ is the number of partitions of $N \in \NN$, and, 

\begin{equation}
\cH \ll \mathbf{\Delta} \, | \, q \rr = 
\sum_{N = 0}^{\, \infty} H_{\, N} \ll \, \alpha_{\, ext}, \Delta_{\, int} \rr \, q^{\, N}
\label{H.expansion}
\end{equation}

\subsubsection{Remark} The factor $q^{\, 1/24} / \eta (q)$ is the character of the Fock space of 
a free boson, and in \eqref{onept}, the 1-point function on the torus is written in terms of the 
free-boson of Feigin and Fuks \cite{feigin.fuks.01, feigin.fuks.02}. This will become clear once 
we take the $\alpha_{\, ext} \rightarrow 0$ limit, and 1-point function becomes a character, in 
section \ref{section.05}.  

\subsubsection{The recursion relation} The recursion comes in the definition of $H_N$, 
the coefficients of the numerator of the conformal block, 

\begin{multline}
H_N \ll \alpha_{\, ext}, \Delta_{\, int} \rr = 
\displaystyle\sum_{rs = 1}^N
\frac{
R^{\, \textit{tor}}_{\, r, s} \ll \alpha_{\, ext} \rr
}{
\ll \Delta_{\, int} - \Delta_{\, r, s} \rr
}
H_{N - r s} \ll \alpha_{\, ext}, \Delta_{\, - r, s} \rr,
H_0 \ll \alpha_{\, ext}, \Delta_{\, int} \rr = 1
\end{multline}

\subsubsection{The  $R^{\, \textit{tor}}_{\, r, s}$ numerators}

\begin{multline}
R^{\, \textit{tor}}_{r,s} \ll \alpha_{\, ext} \rr = 
\frac{1}{4r_{r,s}} 
\prod_{k} 
\prod_{l}
\ll \frac{k-1}{2} \, b + \frac{l-1}{2} \, b^{-1} + \alpha_{\, ext} \rr,
\\
k=1-2r,3-2r, \cdots, 2r-1, 
\quad 
l=1-2s,3-2s, \cdots, 2s-1, 
\label{Rmn}
\end{multline}

\noindent $r_{r,s}$ is given by formula (\ref{rmn}).

\subsubsection{Examples} The simplest coefficients $H_N$, $N = 1, 2, \cdots$, 
in \eqref{H.expansion} are,

\begin{equation}
H_1 \ll \, \alpha_{\, ext}, \Delta_{\, int} \rr =  
\frac{R^{\, \textit{tor}}_{1,1} \ll \alpha_{\, ext} \rr}{\ll \Delta_{\, int} - \Delta_{1,1} \rr}, 
\end{equation}

\begin{multline}
H_2 \ll \, \alpha_{\, ext}, \Delta_{\, int} \rr =  
\frac{R^{\, \textit{tor}}_{1,1} \ll \alpha_{\, ext} \rr}{\ll \Delta_{\, int} -\Delta_{1,1} \rr} 
H_1 \ll \, \alpha_{\, ext}, \Delta_{\, - 1,1} \rr + 
\\ 
\frac{R^{\, \textit{tor}}_{1,2} \ll \alpha_{\, ext} \rr}{\ll \Delta_{\, int} -\Delta_{1,2} \rr} +
\frac{R^{\, \textit{tor}}_{2,1} \ll \alpha_{\, ext} \rr}{\ll \Delta_{\, int} -\Delta_{2,1} \rr}, 
\end{multline}

\begin{multline}
H_3 \ll \, \alpha_{\, ext}, \Delta_{\, int} \rr =  
\\
\frac{R^{\, \textit{tor}}_{1,1} \ll \alpha_{\, ext} \rr}{\ll \Delta_{\, int} - \Delta_{1,1} \rr}     
H_2 \ll \, \alpha_{\, ext}, \Delta_{- 1, 1} \rr + 
\frac{R^{\, \textit{tor}}_{1,2} \ll \alpha_{\, ext} \rr}{\ll \Delta_{\, int} -\Delta_{1,2} \rr}  
H_1 \ll \, \alpha_{\, ext}, \Delta_{- 1, 2} \rr +
\\
\frac{R^{\, \textit{tor}}_{2,1} \ll \alpha_{\, ext} \rr}{\ll \Delta_{\, int} -\Delta_{2,1} \rr}
H_1 \ll \, \alpha_{\, ext}, \Delta_{- 2, 1} \rr + 
\frac{R^{\, \textit{tor}}_{1,3} \ll \alpha_{\, ext} \rr}{\ll \Delta_{\, int} -\Delta_{1,3} \rr} + 
\frac{R^{\, \textit{tor}}_{3,1} \ll \alpha_{\, ext} \rr}{\ll \Delta_{\, int} -\Delta_{3,1} \rr}
\end{multline}

\noindent and so on. 

\subsection{Remark} Expanding \eqref{onept}, we obtain, 
 
\begin{multline}
\cF \ll \Delta_{\, ext}, \Delta_{\, int} \, | \, q \rr =
\,\ll H_0 \rr \,q^0+
\ll H_0 + H_1 \rr \, q \, +
\ll 2 H_0 + H_1 + H_2 \rr \, q^2 \, + \cdots
\\
 = \sum_{N = 0}^{\, \infty} \sum_{k = 0}^{N} p(N-k)H_k q^N
\label{character}
\end{multline}

\noindent From \eqref{character}, the structure of the conformal block
$\cF \ll \mathbf{\Delta} \, | \, q \rr$ is clear. In particular, if $\Delta_{\, ext} = 0$, 
and $H_N = 0$, for all $N = 1, 2, \cdots$, we recover the character of the Fock space of 
a free boson, which is the character of a generic non-minimal conformal field theory. If 
$\Delta_{\, ext} = 0$, and $H_N = \pm 1$, for appropriate values of $N$, null states and 
their descendants are removed and one obtains the character of an irreducible fully-degenerate 
highest-weight module. This will be discussed in detail in section \ref{section.05}.  

\section{The 0-point functions on the torus: The characters} 
\label{section.05}
\noindent \textit{We discuss the derivation of the character of the representation corresponding to $\Delta_{\, int}$ using the recursion relation and a particular limiting procedure, in three cases: 
\1 general central charge and general $\Delta_{\, int}$ , 
\2 general central charge and degenerate representation $\Delta_{\, int}=\Delta_{m,n}$, and 
\3 minimal models $\mathcal{M}_{p,\pp}$ characters.}

\subsection{General central charge and general $\Delta_{\, int}$} 
We introduce regularization parameter $\eps$ that we set to $0$ at the end. We set the inserted vertex-operator to be the identity, in the limit $\eps$ to zero, 

\begin{equation}
\alpha_{\, ext} = 2\eps
\end{equation}

\noindent The factor $2$ in the above definition is for convenience. For general $r,s \in \mathbb{N}$, the term,

\begin{equation}
R^{\, \textit{tor}}_{\, r, s} \, 
\frac{ 
\ll \alpha_{\, ext}=2 \eps \rr
}{
\ll \Delta_{\, int} - \Delta_{r, s} \rr
} = 
- \eps \, 
\frac{
\ll 
Q - 2 \alpha_{r,s} \rr
}{
\ll \Delta_{\, int} - \Delta_{r, s} \rr} + \cO
\ll \eps^2\rr,
\end{equation}

\noindent vanishes in the limit $\eps \rightarrow 0$, provided that $\lim_{\eps \to 0} \Delta_{\, int} \neq \Delta_{r, s}$. All the  $H_i$ are then zero and the expansion is given by (\ref{dchar}). As expected, one finds the character $\chi_{\Delta_{\, int}}(q)$ of an irreducible Verma module of dimension $\Delta_{\, int}$.

\subsection{General central charge, $\Delta_{\, int}=\Delta_{m,n}$.}
First we set  $\alpha_{\, ext}=2 \eps$. Differently from the previous case,  here we encounter the pole coming from the denominator $\ll \Delta_{\, int}-\Delta_{m, n}\rr$. 

\subsubsection{Internal field regularization.} 
We need to regularize the dimension of the internal field, and we set, 

\begin{equation}
\label{intchargereg}
\alpha_{\, int} = 
\alpha_{m, n} + \eps',
\end{equation}

\noindent and we define,

\begin{equation}
\Delta^{(\eps')}_{m,n} =
\alpha_{\, int} \ll Q - \alpha_{\, int} \rr 
= \Delta_{m,n} + \eps' \ll Q - 2 \alpha_{m, n} \rr+\cO\ll \eps{\, \prime \, 2} \rr
\end{equation}

\noindent For $\eps,\eps' << 1$, the term,

\begin{equation}
\frac{R^{\, \textit{tor}}_{\, m, n} \ll 2\eps \rr}{\ll \Delta^{(\eps')}_{m,n} - \Delta_{m, n} \rr} = - \frac{\eps+\cO\ll \eps^2\rr}{\eps'+\cO \ll \eps^{\, \prime \, 2} \rr}
\end{equation} 

\noindent The result of the limit $(\eps,\eps') \to (0,0)$ depends therefore on the way one reaches the point $(\eps,\eps') = (0,0)$. For instance, if one first sends $\eps \to 0$ and then $\eps'\to 0$, all the $H_i$ are zero and the character of a general Verma module is found. This result can be interpreted by saying that the null-state at level $n m$ is not vanishing. Interestingly, such representation appears for instance in the construction of the Liouville theory for $c\leq 1$ \cite{ribault.santachiara}.  By setting $\eps' = \eps$ one finds instead that,

\begin{equation}
\lim_{\eps \to 0}
\frac{R^{\, \textit{tor}}_{m,n} \ll \alpha_{\, ext}=2\eps \rr}{\ll \Delta^{(\eps)}_{m,n} -\Delta_{m, n} \rr} =  - 1,
\label{R}
\end{equation}
for general $b$. The contribution \eqref{R} 
is the only non-zero term in $H_{\, m n}$, which is itself the only non-zero $H_i$. This contribution 
$H_{\, m n} = - 1$ at level $mn$ corresponds to removing the null state. From equation \eqref{character}, 
you can see that the expansion is (keeping only the non-zero terms), 

\begin{multline}
\cF \ll \mathbf{\Delta} \, | \, q \rr = 
1 + \cdots + 
\ll p_{\, mn    } - 1 \rr \, q^{\, m n    } + 
\ll p_{\, mn + 1} - 1 \rr \, q^{\, mn + 1}  
\\
+ \cdots + 
\ll p_{\, mn + N} - p_{\, N} \rr \, q^{\, mn + N} + \cdots 
\end{multline}
This corresponds to quotienting out the module of the null state. 
\noindent We observe that the condition $\eps'=\eps$, providing the character of a degenerate representation $\cV_{m,n}$, assures that the Coulomb gas fusion condition $\alpha_{\, int}+\alpha_{\, int}=\alpha_{\, ext}$ is satisfied for any value of $\eps$. 

\subsection{Characters in minimal models.} 
In the case of the minimal models, $b^2 = - \frac{\pp}{p}$, where $p$ and $\pp$ 
are coprime positive integers, $0 < p < \pp$,  we know that a fully-degenerate highest-weight
module $\cV_{m,n}$ has a null-state at level $mn$, and another at level $(p - m) (\pp - n)$. We need first to solve the new poles appearing in the term 
$H_{\, m^{\, \prime} n^{\, \prime}} = 
\frac{
R^{\, \textit{tor}}_{\, m^{\, \prime}, n^{\, \prime}} \ll 2\eps \rr
}{
\ll \Delta^{(\eps')}_{m,n} - \Delta_{\, m^{\, \prime}, n^{\, \prime}} \rr
}$, where $\Delta^{(\eps')}_{m,n}$ is defined in (\ref{intchargereg}), $0 < m < p$ and $0 < n < \pp$, and  $m^{\, \prime} = p - m$, and $n^{\, \prime} = \pp - n$. 

\subsubsection{The regularization $\eps^{\, \prime \prime}$}
We introduce a third regularization parameter $\eps^{''}$ to move away from the minimal model point by setting

\begin{equation}
b = \sqrt{-\frac{\pp}{p} \ll 1 + \eps^{''} \rr}
\end{equation}
 
\noindent For $\eps,\eps',\eps''<< 1$, 
 
\begin{equation}
\frac{
R^{\, \textit{tor}}_{\, m^{\, \prime}, n^{\, \prime}} \ll 2\eps \rr
}{
\ll \Delta^{(\eps')}_{m,n} - \Delta_{\, m^{\, \prime}, n^{\, \prime}} \rr} = 
- \frac{
\eps + \cO \ll \eps^2 \rr
}{
- \eps' + \frac12 \, \sqrt{-p \pp} \, \eps^{''} + 
\cO 
\ll 
\eps^{\, \prime        \, 2}, 
\eps^{\, \prime \prime \, 2},
\eps' \, 
\eps^{\prime \prime} 
\rr
}
\end{equation}

\noindent Again, the final result depends on how we approach the point $(\eps,\eps',\eps^{''})= (0,0,0)$. In order to obtain the minimal model character we have first to remove the null state at level $(p - m) (\pp - n)$. This is obtained by setting,  

\begin{equation}
\eps^{\, \prime}        \rightarrow \eps,
\quad
\eps^{\, \prime \prime} \rightarrow 
\frac{4}{\sqrt{- \, p \, \pp}} \eps, 
\label{totalreg}
\end{equation}

\noindent and taking the limit $\eps\to 0$. Then both (\ref{R}) and,

\begin{equation}
\lim_{\eps \to 0} \frac{
 R^{\, \textit{tor}}_{\, m^{\, \prime}, n^{\, \prime}} \ll 2\eps \rr
}{
\ll \Delta^{(\eps)}_{m,n} - \Delta_{\, m^{\, \prime}, n^{\, \prime}} \rr
} =  - 1, 
\label{R2}
\end{equation}

\noindent are satisfied at the same time. The results (\ref{R}) and (\ref{R2}) are not sufficient to prove that one obtains in the limit the minimal model character. One has to consider that there are other terms, of the form,

\begin{equation}
\frac{R^{\, \textit{tor}}_{m_1, n_1} \ll 2\eps \rr}{\ll \Delta^{(\eps)}_{m_1,n_1}        - \Delta_{m_1, n_1} \rr}
\frac{R^{\, \textit{tor}}_{m_2, n_2} \ll 2\eps \rr}{\ll \Delta_{- m_1, n_1} - \Delta_{m_2, n_2} \rr}, 
\end{equation} 

\noindent that will contribute when "resonances" in the conformal dimension, such as $\Delta_{- m_1, n_1} = \Delta_{m_2, n_2}$, occur. The fact that there is an infinite number of resonances, or equivalently, infinitely-many pairs $\ll r, s \rr$ 
that correspond to the same conformal dimension $\Delta_{\, r, s}$, and that the recursion relation expression 
for $H_N$ includes all resonances $\ll r, s \rr$ such that $rs \leq N$ will thus play an important role. We are now in the position to give the exact contribution of all the terms which do not vanish in the limit $\eps \to 0$ (see appendix \ref{appendix.b} for the derivation of the following results). In the following we always assume that the integers $\ll m,n\rr$  belong to the minimal model $\cM_{p,\pp}$ Kac table,   $0<\; m\; <\;p, \; 0< \; n \;< \;\pp$. Concerning the terms in the recursion that have at the denominator the dimension of the internal field $\Delta^{(\eps)}_{m,n}$, we show in appendix \ref{appendix.b} that,
 
\begin{multline}
\text{if} \ll m, n \rr \leftrightarrow \ll r, s \rr^{\pm}_{l} \implies 
\lim_{\eps \to 0}\frac{R^{\, \textit{tor}}_{r,s} \ll 2\eps \rr}{\Delta^{(\eps)}_{m,n}-\Delta_{r,s}}=-\frac{1}{2^{2l-1\pm1}(2l\pm1)}\prod_{k = 1}^{l-\frac{1}{2}\pm \frac12} \frac{4k^2-1}{k^2}
\label{firstypeterm}
\end{multline}

\noindent We consider now the terms of the type 
$R^{\, \textit{tor}}_{r,s}\ll 2 \eps \rr/ 
\ll \Delta_{m', - n'} - \Delta_{r, s} \rr$. 
We define two integers $(l_1,l_2)$ from the Euclidean division of $\ll r,s \rr$, $\ll r,s\rr =\ll l_1 p+ m,\; l_2 \pp+n\rr $. We have,

\begin{multline}
\text{if} \ll m', -n' \rr \leftrightarrow \ll r, s \rr^{\pm}_{l'}
\implies 
\\
\lim_{\eps \to 0}\frac{R^{\, \textit{tor}}_{r,s}}{\Delta_{m',-n'}-\Delta_{r,s}}=-\frac{1}{2^{2\; l+1} \; l'}\;\;\prod_{k = 1}^{l}\frac{4k^2-1}{k^2},\quad l=\text{min}\ll l_1,l_2\rr
\label{secondtypeterm}
\end{multline}

\noindent We will provide below the complete combinatorial structure  of all the terms that contribute to the character. All these terms are finite, but they are fractional and add up to integral values.

\subsection{Example 1. The Ising model, 
$( \pp, p) = (4, 3)$, 
$( m,   n) = (1, 1)$
}
We give here an explicit application of the previous formulas to the identity character of the  $\cM_{4,3}$ minimal model. We set $\alpha_{est}=2\eps$,  $\alpha_{int}=\eps$ and $b = b^\eps$ given by (\ref{totalreg}). 

\begin{multline}
H_{\, 1} = 
\frac{R^{\, \textit{tor}}_{1, 1} \ll 2 \eps \rr}{\ll \Delta^{(\eps)}_{1,1}  - \Delta_{1, 1} \rr}, 
\quad 
\lim_{\eps\to 0} H_{\, 1}=-1, 
\quad 
H_{\, 6} = 
\frac{R^{\, \textit{tor}}_{2, 3} \ll 2 \eps \rr}{\ll \Delta^{(\eps)}_{1,1}  - \Delta_{2, 3} \rr}, 
\quad 
\lim_{\eps\to 0} H_{\, 6}=-1
\\
H_{\, 11} = 
\frac{R^{\, \textit{tor}}_{1,1} \ll 2 \eps \rr}{\ll \Delta^{(\eps)}_{1,1}   - \Delta_{1,1} \rr}
\frac{R^{\, \textit{tor}}_{2, 5} \ll 2 \eps \rr}{\ll \Delta_{1, -1} - \Delta_{2, 5} \rr} 
+ 
\frac{R^{\, \textit{tor}}_{2, 3} \ll 2 \eps \rr }{\ll \Delta^{(\eps)}_{1,1}   - \Delta_{2, 3} \rr}
\frac{R^{\, \textit{tor}}_{5, 1} \ll 2 \eps \rr }{\ll \Delta_{2, -3} - \Delta_{5, 1} \rr}
\end{multline}

\noindent Using the fact that  $\ll 1,1\rr$ and $\ll 2, 3\rr$ are in the Kac table and that $\ll 1,-1 \rr \rightarrow \ll 2, 5 \rr^{-}_{1}$, $\ll 2, -3 \rr \rightarrow \ll 5, 1 \rr^{+}_{1}$, one has from (\ref{firstypeterm}) and (\ref{secondtypeterm}),

\begin{equation}
\lim_{\eps \to 0} H_{\, 11} = \ll -1 \times -\frac12\rr +\ll -1 \times -\frac12\rr = 1
\end{equation}

\begin{equation}
H_{\, 13} = 
\frac{R^{\, \textit{tor}}_{1,1} \ll 2 \eps \rr}{\ll \Delta^{(\eps)}_{1,1} - \Delta_{1,1} \rr}
\frac{R^{\, \textit{tor}}_{4,3} \ll 2 \eps \rr}{\ll \Delta_{-1, 1} - \Delta_{4, 3} \rr} 
+ 
\frac{R^{\, \textit{tor}}_{2, 3} \ll 2 \eps\rr}{\ll \Delta^{(\eps)}_{1,1}  - \Delta_{2,3} \rr}
\frac{R^{\ \textit{tor}}_{1, 7} \ll 2 \eps \rr}{\ll \Delta_{2, -3} - \Delta_{1, 7} \rr}
\end{equation}
From $\ll 1,-1 \rr \rightarrow \ll 4, 3 \rr^{+}_{1}$, $\ll 2, -3 \rr \rightarrow \ll 1, 7 \rr^{-}_{1}$
\begin{equation}
\lim_{\eps \to 0} H_{\, 13} = \ll -1 \times -\frac12\rr +\ll -1 \times -\frac12\rr = 1,  
\end{equation}

\noindent These two terms correspond to adding again the null states at level $11$ and $13$, which are contained into the modules of both the null states at level $1$ and $6$, and were therefore subtracted twice.
Let us make another example at level $20$,

\begin{multline}
H_{\, 20} = 
\frac{ R^{\, \textit{tor}}_{1,1} \ll 2 \eps \rr}{ \ll\Delta^{(\eps)}_{1,1}  - \Delta_{1, 1} \rr}
\frac{ R^{\, \textit{tor}}_{4,3} \ll 2 \eps \rr}{ \ll \Delta_{1, -1} - \Delta_{4, 3} \rr}
\frac{ R^{\, \textit{tor}}_{7,1} \ll 2 \eps \rr}{ \ll \Delta_{4, -3} - \Delta_{7, 1} \rr} 
+ 
\\
\frac{R^{\, \textit{tor}}_{1,1} \ll 2 \eps\rr}{\ll \Delta^{(\eps)}_{1,1}   - \Delta_{1, 1} \rr}
\frac{R^{\, \textit{tor}}_{2,5} \ll 2 \eps \rr}{\ll \Delta_{1, -1} - \Delta_{2, 5} \rr}
\frac{R^{\, \textit{tor}}_{1,9} \ll 2 \eps \rr}{\ll \Delta_{2, -5} - \Delta_{1, 9} \rr} 
+ 
\\ 
\frac{R^{\, \textit{tor}}_{2,3} \ll 2 \eps \rr}{\ll \Delta^{(\eps)}_{1,1}   - \Delta_{2, 3} \rr}
\frac{R^{\, \textit{tor}}_{5,1} \ll 2 \eps \rr}{\ll \Delta_{2, -3} - \Delta_{5, 1} \rr}
\frac{R^{\, \textit{tor}}_{1,9} \ll 2 \eps \rr}{\ll \Delta_{5, -1} - \Delta_{1, 9} \rr}
+ 
\\
\frac{R^{\, \textit{tor}}_{2, 3} \ll 2 \eps\rr}{\ll \Delta^{(\eps)}_{1,1}   - \Delta_{2, 3} \rr} 
\frac{R^{\, \textit{tor}}_{1, 7} \ll 2 \eps \rr}{\ll \Delta_{2, -3} - \Delta_{1, 7} \rr}
\frac{R^{\, \textit{tor}}_{7, 1} \ll 2 \eps \rr}{\ll \Delta_{1, -7} - \Delta_{7, 1} \rr}
+ 
\frac{R^{\, \textit{tor}}_{4, 5} \ll 2 \eps \rr}{\ll \Delta^{(\eps)}_{1,1}  -  \Delta_{4, 5} \rr}
\label{ising20c}
\end{multline}

\noindent From (\ref{firstypeterm}) and (\ref{secondtypeterm}),
\begin{multline}
\lim_{\eps \to 0}  H_{\, 20} = 
\ll -1 \times - \frac12 \times - \frac12 \rr + 
\ll -1 \times - \frac12 \times - \frac12 \rr + 
\\
\ll -1 \times - \frac12 \times - \frac14 \rr + 
\ll -1 \times - \frac12 \times - \frac14 \rr - 
\frac14 
= -1
\end{multline}

\noindent Notice that this sum of terms which add up to an integer has exactly the same structure as equation (\ref{ising20}).

\subsection{General case}

We consider the character of a representation indexed by $\ll m,n \rr$, $0< m < p$, $0 < n < \pp$.  We provide here the explicit procedure to find all the terms that, in the limit (\ref{totalreg}), have a finite fraction contribution that sums up, at a given level, to $1$ or $-1$. 
We want to give a procedure that takes into account  all the singular terms appearing in the one-point torus recursion relation in the minimal model limit.  
Given a pair of indices $\ll m,n \rr$, we want to find the set of pairs $\ll r,s\rr$, that are resonant with $\ll m,-n\rr$, $\ll m,-n\rr \leftrightarrow \ll r,s\rr^{+}_{l}$, or with $\ll -m,n\rr$, $\ll -m,n\rr \leftrightarrow \ll r,s\rr^{+}_{l}$, where we used the notation defined in (\ref{resnotation}). These pairs are obtained respectively by the two  transformations, 

\begin{align}
& v^{(l)}_1: \quad \ll m,n\rr \rightarrow \ll r= l p+ m, s= l \pp - n \rr \quad \ll m,-n\rr \leftrightarrow \ll r,s\rr^{+}_{l} 
\\
& v^{(l)}_2: \quad \ll m,n\rr \rightarrow \ll r=l p- m, s=l \pp + n \rr 
\quad 
\ll -m,n\rr \leftrightarrow \ll r,s\rr^{+}_{l} 
\end{align}
By using these transformations we generate the following diagram,

 \begin{tikzpicture}[scale=1,every node/.style={scale=1}]
 \draw(-0.75,3) node[left]{$k=0$};
 \draw(0,0) node[left]{$\ll m,  n \rr$};
 \draw[decoration={brace,mirror},decorate]
  (-3.5,1) --node[left]{$\Delta_{m,n}$}(-3.5,-4);
 \draw(0.65,-1) node[left]{$\ll p+m,  \pp +n \rr$};
 \draw(0.65,-2) node[left]{$\ll 2p+m,  2\pp +n \rr$};
 \draw(0,-3) node[left]{$\cdots \cdots$};
 \draw[thick,->](0,0)--(2.15,0.5);
 \draw(1,0.5) node[above]{$v^{(l)}_{1}$};
 \draw[thick,->](0,0)--(2.15,-2);
 \draw(1,-2) node[above]{$v^{(l)}_{2}$};
 \draw[decoration={brace,mirror},decorate]
  (-3.5,-5) --node[left]{$\Delta_{m,n}$}(-3.5,-9);
 \draw(0.5,-6) node[left]{$ \ll p-m, \pp-n \rr$};
 \draw(0.5,-7) node[left]{$ \ll 2p-m, 2\pp-n \rr$};
 \draw(0,-8) node[left]{$\cdots \cdots$};
 \draw[thick,->](0.5,-6)--(2.15,-5.5);
 \draw(1,-5.5) node[above]{$v^{(l)}_{1}$};
 \draw[thick,->](0.5,-6)--(2.15,-8);
 \draw(1,-8) node[above]{$v^{(l)}_{2}$};
 \draw(5,3) node[left]{$k=1$};
 \draw[decoration={brace},decorate]
  (6,2) --node[right]{$\Delta_{m,n}+ m n$}(6,-1);
 \draw(6,1) node[left]{$\ll p+ m, \pp- n \rr$};
 \draw(8,1) node[above]{$v^{(l)}_{1}$};
 \draw[thick,->](5.75,1)--(12,1);
 \draw(6,0) node[left]{$\ll 2 p+m,  2 \pp -n \rr$};
 \draw(5,-1) node[left]{$\cdots \cdots$};
 \draw[decoration={brace},decorate]
  (6,-1.1) --node[right]{$\Delta_{m,n}+ m n$}(6,-4.15);
 \draw(6,-2) node[left]{$\ll p-m,  \pp +n \rr$};
 \draw(6,-3) node[left]{$\ll 2p-m,  2\pp +n \rr$};
 \draw(6,-4) node[left]{$\cdots \cdots$};
 \draw[decoration={brace},decorate]
  (6,-4.3) --node[right]{$\Delta_{m,n}+ (p-m)(\pp-n)$}(6,-7);
 \draw(6,-5) node[left]{$ \ll m, 2\pp-n \rr$};
  \draw(7,-4) node[above]{$v^{(l)}_{1}$};
 \draw[thick,->](6,-5)--(12,0.8);
 \draw(9,-4) node[below]{$v^{(l)}_{2}$};
 \draw[thick,->](6,-5)--(12,-3);
 \draw(6,-6) node[left]{$ \ll p+m, 3\pp-n \rr$};
 \draw(6,-7) node[left]{$\cdots \cdots$};
 \draw(6,-8) node[left]{$ \ll 2 p-m, n \rr$};
 \draw(7.5,-7.2) node[above]{$v^{(l)}_{1}$};
 \draw[thick,->](5.75,-8)--(12,-8);
 \draw(7.5,-8) node[above]{$v^{(l)}_{2}$};
 \draw[thick,->](5.75,-8)--(11,-5.75);
 \draw(6,-9) node[left]{$ \ll 3p-m, \pp+n \rr$};
 \draw(6,-10) node[left]{$\cdots \cdots$};
  \draw[decoration={brace},decorate]
  (6,-7.1) --node[right]{$\Delta_{m,n}+ (p-m)(\pp-n)$}(6,-11);
  \draw(13,3) node[left]{$k=2$};
  \draw(6,-10) node[left]{$\cdots \cdots$};
  \draw[decoration={brace},decorate]
  (13,-7.1) --node[left]{$\cdots$}(13,-11);
  \draw[decoration={brace},decorate]
  (13,-4.3) --node[left]{$\cdots$}(13,-7);
  \draw[decoration={brace},decorate]
  (13,-1.1) --node[left]{$\cdots$}(13,-4.15);
  \draw[decoration={brace},decorate]
  (13,2) --node[left]{$\cdots$}(13,-1);
 \end{tikzpicture}
 
\noindent In the $k=0$ column, we place the two groups of pairs, $\ll r,s\rr$, which are resonant with $\ll m,n\rr$,  $\ll m,n\rr \leftrightarrow \ll r,s\rr^{+}_l$ and $\ll m,n\rr \leftrightarrow \ll r,s\rr^{-}_l$. The column $k=1$ is generated by applying the transformations $v^{(l)}_{1}$ and $v^{(l)}_{2}$ to $\ll m,n\rr$ and $\ll \pp-m,p-n\rr$. One obtains four families, corresponding to the two set of pairs $\ll r,s \rr$  resonant with $\ll m,-n \rr$, $\ll m,-n\rr \leftrightarrow \ll r,s\rr^{+}_l$ and $\ll m,-n\rr \leftrightarrow \ll r,s\rr^{-}_l$ and associated to representation with dimension $\Delta_{m,n}+m n$,  plus the two set of pairs in resonance with $\ll p-m, n-\pp\rr$, $\ll p-m,n-\pp\rr \leftrightarrow \ll r,s\rr^{+}_l$ and $\ll p-m,n-\pp\rr \leftrightarrow \ll r,s\rr^{-}_l$ and associated to representation with dimension $\Delta_{m,n}+(p'-m)(p-n)$. The column $k=2$ is obtained by applying the transformations $v^{(l)}_{1}$ and $v^{(l)}_{2}$ to $\ll p+m,\pp-n\rr$,  $\ll p-m,\pp+n\rr$, $\ll m,2\pp-n\rr$ and $\ll 2p-m,n\rr$ and so on. At each column one can therefore identify four families of pairs that we indicate with the letters $U^{1}, D^{1}$ and $U^{2}, D^{2}$. Any pair of indices appearing in the diagram is identified by its family and by two non-negative  integers $k$ and $l$, indicating respectively the column and the position in the interior of each family, 

\begin{multline}
U^{1}(k,l) \rightarrow  \ll (k+l)p+m,(-1)^k n+ l \pp + \frac{1-(-1)^k}{2} \pp\rr 
\\
D^{1}(k,l) \rightarrow  \ll (1+l)p-m,(-1)^{k+1} n+ (l + k) \pp + \frac{1-(-1)^{k+1}}{2} \pp\rr 
\\
U^{2}(k,l) \rightarrow  \ll l p+m,(-1)^k n+ (l+k) \pp + \frac{1-(-1)^k}{2} \pp\rr 
\\
D^{2}(k,l) \rightarrow  \ll (1+l+k)p-m,(-1)^{k+1} n+ l \pp + \frac{1-(-1)^{k+1}}{2} \pp\rr
\end{multline}

\noindent It is straightforward to obtain the action of the transformations $v^{(l)}_{1}$ and $v^{(l)}_{2}$ on each index,

\begin{multline}
v^{(l')}_{1}: U^{1}(k,l)\to  U^{1}(k+2l+1,l'), \quad U^{2}(k,l)\to  U^{1}(k+2l+1,l') \\
v^{(l')}_{1}: D^{1}(k,l)\to  D^{2}(k+2l+1,l'), \quad D^{2}(k,l)\to  D^{2}(k+2l+1,l')\\
v^{(l')}_{2}: U^{1}(k,l)\to  D^{1}(k+2l+1,l'), \quad U^{2}(k,l)\to  D^{1}(k+2l+1,l') \\
v^{(l')}_{2}: D^{1}(k,l)\to  U^{2}(k+2l+1,l'), \quad D^{2}(k,l)\to  U^{2}(k+2l+1,l').
\label{transformations}
\end{multline}

\noindent The above rules allow to write the chains of resonant terms in the recursion relation in terms of words formed by the letters $U^{1,2}$ and $D^{1,2}$, whose sequences have to satisfy the connections above.
For instance, the terms in the example in (\ref{ising20c}), correspond to the following words, 

\begin{align}
&U^{1}(0,0)U^{1}(1,0)U^{1}(2,0)\nonumber\\
&U^{1}(0,0)D^{1}(1,0)U^{2}(2,0)\nonumber\\
&D^{1}(0,0)D^{2}(1,0)U^{2}(2,0)\nonumber \\
&D^{1}(0,0)U^{2}(1,0)U^{1}(2,0)\nonumber\\
&U^{1}(0,1)\nonumber
\end{align}

\noindent As seen in this example, the words corresponding to a certain level $N$ will all end either with $U$ or with $D$. The reason is that the last arrows of the chains must point to the same dimension, so they must all point either to the $1$ sector or all point to the $2$ sector. From (\ref{transformations}), all $U$ labels transform to a label in the $1$ sector, while all $D$ labels transform to labels in the $2$ sector.
Since there always exists either a $U^1(0,0)U^1(1,0) \cdots U^1(K,0)$ or a $D^2(0,0)D^2(1,0)\cdots D^2(K,0)$ chain, of length $K+1$, the level $N$ is either,

\begin{equation}
N_U(K) = 
\sum_{k=0}^K 
\ll k p + m \rr 
\ll 
(-1)^k n + \ll 1 - (-1)^k  \rr 
\frac{\pp}{2}
\rr, 
\end{equation}

\noindent or

\begin{equation}
N_D(K) = 
\sum_{k=0}^K 
\ll (k+1) p - m \rr 
\ll (-1)^{k+1} n + \ll 1 - (-1)^{k+1} \rr 
\frac{\pp}{2}
\rr 
\end{equation}

\noindent Then, the  contribution of each  each word is obtained by using formulas (\ref{firstypeterm}) and (\ref{secondtypeterm}).
For instance,

\begin{multline}
U^{1} (k_1,l_1) U^{1} (k_1+2 l_1+1,l_2)
\rightarrow 
\lim_{\epsilon\to 0}
\frac{
R^{\, \textit{tor}}_{r_2,s_2} (2\epsilon)
}{
\Delta_{r_1,-s_1}-\Delta_{r_2,s_2}
}, 
\\
\textit{with} \quad 
\ll r_1, s_1 \rr = U^{1} (k_1,l_1), \quad 
\ll r_2, s_2 \rr = U^{1}(k_1+2 l_1 + 1, l_2)
\end{multline}

\noindent Applying (\ref{secondtypeterm}) with $l'=l_1+l_2+1$ and $l=l_2$, one has,

\begin{equation}
U^{1}(k_1,l_1)U^{1}(k_1+2 l_1+1,l_2)\rightarrow -\frac{1}{2 ^{2l_{2}+1} (l_1+l_2+1)}\prod_{j=1}^{l_2} \frac{4 \lambda^2-1}{\lambda^2}
\end{equation}
In the same way, we find,
\begin{multline}
U^{2}(k_1,l_1)D^{1}(k_1+2l_1+1,l_2) = D^{1}(k_1,l_1)U^2(k_1+2l_1+1,l_2)\\= D^2(k_1,l_1)D^2(k_1+2l_1+1,l_2)=U^1(k_1,l_1)U^1(k_1+2l_1+1)\\=-\frac{1}{2^{2l_2+1}(l_1+l_2+1)}\prod_{\lambda=1}^{l_2}\frac{4\lambda^2-1}{\lambda^2}
\label{termsa}
\end{multline}
and
\begin{multline}
U^2(k_1,l_1)U^1(k_1+2l_1+1) = D^1(k_1,l_1)D^2(k_1+2l_1+1,l_2)\\
=D^2(k_1,l_1)U^2(k_1+2l_1+1,l_2)=U^1(k_1,l_1)D^1(k_1+2l_1+1,l_2)\\=-\frac{1}{2^{2l_2+1}(k_1+l_1+l_2+1)}\prod_{\lambda=1}^{l_2}\frac{4\lambda^2-1}{\lambda^2}
\label{termsb}
\end{multline}
The non-trivial contribution at level $N(K)$ is given by all possible chains starting with $U^1(0,l_0)$ or $D^1(0,l_0)$, with constraint on the last terms : $U^1$ or $U^2$ if $N=N_1$, $D^1$ or $D^2$ if $N=N_2$. Note that given equalities (\ref{termsa}) and (\ref{termsb}), for fixed $K$ and fixed $\{l_i\}$, the contributions at levels $N_1(K)$ and $N_2(K)$ are equal. Therefore, to compute the contribution at level $N_1(K)$ (resp. at level $N_2(K)$), instead of constraining the chains to end by $U$ (resp. $D$), we can leave the ends of the chains free and divide by $2$ at the end.
The first terms of the chains involve the internal dimension $\Delta^\eps_{m,n}$,

\begin{multline}
U^1(0,l_0) = 
\frac{
R^{\, \textit{tor}}_{l_0 p+m,l_0 \pp+n}
}{
\Delta^\eps_{m,n}-\Delta_{l_0 p+m,l_0 \pp+n}
} = 
- 
\frac{
1
}{
2^{2l_0}(2l_0+1)
}
\prod_{\lambda=1}^{l_0}
\frac{
4 \lambda^2 - 1
}{
\lambda^2
}, 
\\
D^1(0,l_0) = \frac{R^{\, \textit{tor}}_{(l_0+1)p-m,(l_0+1)\pp-n}}{\Delta^\eps_{m,n}-\Delta_{(l_0+1)p-m,(l_0+1)\pp-n}} = -\frac{1}{2^{2l_0}(2l_0+1)}\prod_{\lambda=1}^{l_0}\frac{4\lambda^2-1}{\lambda^2}
\label{term0}
\end{multline}

\noindent We now need to specify the set $\{l_i\}$. Let us denote $I$ the cardinal of this set. We then have the constraint,

\begin{equation}
I+\sum_{i=0}^{I-1}2l_i=K+1
\end{equation}
By writing $I = K+1-2a$, the constraint is,

\begin{equation}
\sum_{i=0}^{K-2a}l_i=a, 
\end{equation}

\noindent and the set $\{l_0,l_1, \cdots,$ $l_{K-2a}\}$ is therefore a partition of size $K-2a+1$ of the integer $a$, 
$\{l_0, l_1, \cdots, $ $l_{K-2a} \} = \tilde{p}_{K-2a+1}(a)$ 
in which zeroes are included, as well as all permutations. 
For example the set of the partitions of size $3$ of $3$ is,

\begin{multline}
\{(3,0,0), (0,3,0), (2,1,0), (2,0,1), (1,2,0),
(1,0,2), (0,1,2), (0,2,1), (1,1,1)\}. 
\end{multline}

\noindent $a$ then runs from $0$ to a maximal $a$. If K is odd, $K+1$ is even and $I_{min}=K+1-2a_{max}$ is even. Therefore $I_{min}=2$, and $a_{max}=(K-1)/2$. If $K$ is even, $K+1$ is odd and $I_{min}=K+1-2a_{max}$ is odd and equals 1. Therefore $a_{max}=K/2$ and $l_0=a_{max}$, all other $l_i$ being $0$, and this chain consists of the single term (\ref{term0}). We can now give the final formula for the contribution $C_{N(K)}$ that takes into account all the non-zero terms at a given level $N(K)$,

\begin{multline}
C_{N(K)} = \frac12 \,  
\sum_{a=0}^{[\frac{K-1}{2}]}
\sum_{\{l_0, \cdots, l_{K-2a} \} = \tilde{p}_{K-2a+1}(a)} 
\sum_{\{X_0, \cdots, X_{K-2a}\}} 
X_0 \ll 0, l_0 \rr 
\\ 
\prod_{j=0}^{K-1-2a}
X_{j  } \ll 2 \sum_{i=0}^{j-1} l_i + j,  l_{j  } \rr 
X_{j+1} \ll 2 \sum_{i=0}^{j  } l_i + j+1,l_{j+1} \rr 
\\
+ \delta_{K(\text{mod }2),0} X_0 \ll 0, K/2 \rr, 
\end{multline}

\noindent where the $X_{j>1}$'s are $U^{1,2}$ or $D^{1,2}$, $X_0$ is $U^1$ or $D^1$. From (\ref{termsa}), (\ref{termsb}) and (\ref{term0}),

\begin{multline}
C_{N(K)} = - \delta_{K(\text{mod }2),0}\frac{\displaystyle\prod_{\lambda=1}^{K/2}\frac{4\lambda^2-1}{\lambda^2}}{2^{K}(K+1)} + \sum_{a=0}^{[\frac{K-1}{2}]}\sum_{\{l_0,..,l_{K-2a}\}=\tilde{p}_{K-2a+1}(a)}-\frac{\displaystyle\prod_{\lambda=1}^{l_0}\frac{4\lambda^2-1}{\lambda^2}}{2^{2l_0}(2l_0+1)}
\\
\times\prod_{j=0}^{K-1-2a}-\frac{\displaystyle\prod_{\lambda=1}^{l_{j+1}}\frac{4\lambda^2-1}{\lambda^2}}{2^{2l_{j+1}+1}}\ll \frac{1}{l_j+l_{j+1}+1}+\frac{1}{l_j+l_{j+1}+1+2\sum_{i=0}^{j-1}l_i +j}\rr .
\end{multline}

\noindent We have checked numerically - up to order $N \sim 300$ - that $C_{N(K)} = (-1)^{K+1}$.

\section{Conclusions}
We extended Zamolodchikov's elliptic recursion relation for 4-point conformal blocks on 
the sphere \cite{zamolodchikov.02}, and its analogue for 1-point functions on the torus
\cite{fateev.litvinov,poghossian.01}, originally derived for conformal blocks in Liouville 
theory with non-rational central charge, to conformal field theories with rational central 
charges, including the generalized minimal and minimal models. 
When the central charge is rational, solutions of the recursion relation have additional poles 
that appear on a term by term basis. These poles are non physical in the sense that they are 
artifacts of the recursion which splits perfectly well-defined terms into terms that can be 
singular on their own but add up to finite contributions. 
We studied the structure of these non physical poles in two situations. 
\1 In 4-point conformal blocks on the sphere, where we found that the singular terms add up to 
finite terms on the basis of examples, and conjectured that this is the case in general, and that 
regularizing properly all the parameters entering the conformal block, one obtains the minimal 
model conformal block.
\2 In 1-point conformal blocks on the torus, in the limit where the vertex operator insertion 
is the identity operator and the 1-point conformal block reduces to a 0-point conformal block, 
which is a Virasoro character. In this case, the contributions of the non-physical poles are 
fractions, and explicit expressions of these fractions were derived in (\ref{firstypeterm}) 
and (\ref{secondtypeterm}). 
We unveiled the combinatorial structure of these fractions found it to be reminiscent of that 
in the Feigin-Fuks construction of minimal model characters \cite{feigin.fuks.01}, and used it 
to show that the contribution of the non-physical poles add up to $\pm1$. 
The non-physical poles of the 4-point conformal blocks also follow this combinatorial structure. 
A fine regularization of the central charge is needed in the case of the 0-point functions, 
whereas the 4-point function is not sensitive to the regularization used.

\section*{Acknowledgements}
We thank Kenji Iohara, Santiago Migliaccio, Rubik Poghossian and Sylvain Ribault for discussions. 
OF wishes to thank Profs K Lechner, M Matone and D Sorokin for hospitality 
in the Physics Department, University of Padova, while this work was in progress, 
and the organizers of  
\textit{\lq Supersymmetric Quantum Field Theories in the Non-perturbative Regime\rq}, 
and Prof A Dabholkar
for hospitality at the Galileo Galilei Institute for Theoretical Physics, Arcetri,
Firenze, Italy, and at the Abdus Salam Center for Theoretical Physics, Trieste, 
respectively, where it was finalized.
OF is supported by a Special Studies Program grant from the Faculty of Science, 
University of Melbourne, and the Australian Research Council.

\appendix{}

\section{A direct computation at $c=-2$}
\label{appendix.a}
\noindent At $c = -2$, which we can consider as the $\mathcal{M}(2,1)$ minimal model, the first 
extra pole appear at order $3$ in the expansion of the conformal block. It is therefore
possible to compare the recursion result to the result one gets by hand, ie by computing 
the Shapovalov matrix of inner products and the "matrix elements",

\begin{multline}
\langle L_{-Y}\Delta | \Phi_{\Delta_1}(x)\Phi_{\Delta_2}(0)\rangle / \langle \Delta | \Phi_{\Delta_1}(x)\Phi_{\Delta_2}(0)\rangle, 
\\
\langle \Phi_{\Delta_3}(1)\Phi_{\Delta_4}(\infty) |L_{-Y^{\,\prime}} \Delta \rangle / \langle \Phi_{\Delta_3}(1)\Phi_{\Delta_4}(\infty) | \Delta \rangle
\end{multline}

\noindent which appear in (\ref{cbexp}). In the basis 
$\{L_{-1}^3\ket{\Delta},L_{-1}\ket{Q_2},\ket{Q_3}\}$ where,

\begin{multline}
\ket{Q_2}=\ll L_{-1}^2-\frac{2(2\Delta+1)}{3}L_{-2} \rr \ket{\Delta},
\\
\ket{Q_3}=\ll L_{-1}^3-2(\Delta+1)L_{-2}L_{-1}+\Delta(\Delta+1)L_{-3} \rr \ket{\Delta}
\end{multline}

\noindent are the quasi-primary states at levels $2$ and $3$, the Shapovalov matrix at level $3$ is diagonal, 

\begin{multline}
S_{(3)} = 
\text{diag}
\ll  
24\Delta(\Delta+1)(2\Delta+1), 
\right.
\\
\frac{64}{9}(\Delta+2)(2\Delta+1)(\Delta-\Delta_{1,2})(\Delta-\Delta_{2,1}),
\\
\left. 
6\Delta(\Delta+1)(\Delta+2)(\Delta-\Delta_{1,3})(\Delta-\Delta_{3,1}) 
\rr 
\end{multline}

\noindent Then the contribution of the quasi-primary at level $3$ is,

\begin{equation}
\frac{P_{L} \ll Q_3 \, ;\, \Delta_1,\Delta_2 \rr P_{R} \ll Q_3 \, ;\, \Delta_3,\Delta_4 \rr}{\langle Q_3 | Q_3 \rangle} = 
\frac{\mathcal{P}_2 \ll \Delta\, ;\, \Delta_i \rr}{6\Delta(\Delta+1)(\Delta+2)(\Delta-\Delta_{1,3})(\Delta-\Delta_{3,1})}
\label{contrQ3}
\end{equation}

\noindent where $\mathcal{P}_2 \ll \Delta\, ;\, \Delta_i \rr$ is a polynomial of order $2$ in the internal dimension $\Delta$ :

\begin{multline}
\mathcal{P}_2 \ll \Delta\, ;\, \Delta_i \rr = (\Delta_1 - \Delta_2)(\Delta_3 - \Delta_4)(\Delta_1 + \Delta_2 - 1)(\Delta_3 + \Delta_4 - 1)\Delta^2
\\
+ (\Delta_1 - \Delta_2)(\Delta_3 - \Delta_4)\Big\{\ll 1 - (\Delta_1 + \Delta_2)\rr \ll (\Delta_3 - \Delta_4)^2 - (\Delta_3 + \Delta_4)\rr
\\
+ \ll 1 - (\Delta_3 + \Delta_4)\rr \ll (\Delta_1 - \Delta_2)^2 - (\Delta_1 + \Delta_2)\rr \Big\}\Delta
\\
+ (\Delta_1 - \Delta_2)(\Delta_3 - \Delta_4)\ll (\Delta_1-\Delta_2)^2 - (\Delta_1 + \Delta_2)\rr \ll (\Delta_3-\Delta_4)^2 - (\Delta_3 + \Delta_4) \rr
\end{multline}

\noindent When $c = -2$, we have,  

\begin{equation}
\Delta_{1,3} = 0,
\quad 
\Delta_{3,1} = 3, 
\label{dims}
\end{equation}

\noindent and the contribution (\ref{contrQ3}) of $\ket{Q_3}$ is well-defined. However, if we decompose it in partial fractions, 

\begin{equation}
\frac{\langle Q_3 | \Phi_{\Delta_{1}}(0)|\Delta_2\rangle \langle \Delta_3 | \Phi_{\Delta_4}(\infty) | Q_3\rangle}{\langle Q_3 | Q_3 \rangle} =  
\frac{A}{\Delta - \Delta_{1,1}} + \frac{B}{\Delta - \Delta_{1,3}} + \frac{C}{\Delta - \Delta_{3,1}} + \frac{A\Delta + D}{(\Delta + 1)(\Delta + 2)}, 
\end{equation}

\noindent we find,

\begin{multline}
A = \frac{f_{1,1} \ll \Delta_1, \Delta_2 \rr f_{1,1} \ll \Delta_3, \Delta_4 \rr}{2\Delta_{1,3}\Delta_{3,1}},
\\
B = \frac{f_{1,3} \ll \Delta_1, \Delta_2 \rr f_{1,3} \ll \Delta_3, \Delta_4 \rr}{\Delta_{1,3}(\Delta_{1,3}-\Delta_{3,1})(1+\Delta_{1,3})(2+\Delta_{1,3})}, 
\\ 
C = \frac{f_{3,1} \ll \Delta_1, \Delta_2 \rr f_{3,1} \ll \Delta_3, \Delta_4 \rr}{\Delta_{3,1}(\Delta_{1,3}-\Delta_{3,1})(1+\Delta_{3,1})(2+\Delta_{3,1})}, 
\end{multline}

\noindent where we defined the function, 

\begin{equation}
f_{r,s} \ll \Delta_i, \Delta_j \rr = (\Delta_i - \Delta_j) \ll (\Delta_i - \Delta_j)^2 - (\Delta_i + \Delta_j)(1+\Delta_{r,s})+\Delta_{r,s}\rr
\end{equation}

\noindent Notice that given equations (\ref{dims}), $A$ and $B$ become singular for $c = -2$. In fact we have,

\begin{equation}
\lim_{c\to -2} A \propto \frac{R_{1,1}(\Delta_i)R_{2,1}(\Delta_i)}{\Delta_{1, -1} - \Delta_{2, 1}}, 
\quad
\lim_{c\to -2} B \propto R_{1,3}(\Delta_i), 
\quad
\lim_{c\to -2} C \propto R_{3,1}(\Delta_i)
\end{equation}

\noindent $A$ and $B$ are the terms that add to a finite contribution. In that sense, the extra poles are artifacts of the recursion relation rewrites a well-defined quantity as a sum of terms that are individually singular.

\section{About $R^{\, \textit{tor}}_{m,n}$}
\label{appendix.b}

\noindent We prove that 

\begin{equation}
R^{\, \textit{tor}}_{\, m, n}\ll 2\eps\rr= -\eps \ll m b + n b^{-1} \rr + \cO \ll \eps^2 \rr, 
\end{equation}

\noindent for general $b$ and for all $(m,n)$.

\begin{multline}
R^{\, \textit{tor}}_{m,n} \ll \alpha \rr= 
\frac{1}{4r_{m,n}} \prod_{k,l} \ll \frac{1-k}{2} \, b + \frac{1-l}{2} \, b^{-1} - \alpha \rr,
\\
k = 1 - 2m,3 - 2m, \cdots, 2m-1,
\quad 
l = 1 - 2n,3 - 2n, \cdots, 2n-1 
\end{multline}

\noindent and,

\begin{multline}
r_{m,n}=-\frac{1}{2}\prod_{\rho,\sigma} 2\lambda_{\rho, \sigma},\quad \rho=1-m,2-m, \cdots, m, 
\quad 
\sigma=1-n,2-n, \cdots, n, 
\\ 
(\rho,\sigma)\neq (0,0), \;(m,n)
\end{multline}

\noindent which can be rewritten as,  

\begin{equation}
r_{m,n} = \frac{1}{2} \prod_{i=1-m}^m \prod_{j = 1 - n}^n \ll i \, b + j \, b^{-1} \rr, 
\quad
(i,j) \neq (0,0), 
\quad
(i,j) \neq (m,n), 
\end{equation}

\noindent When the inserted operator is the identity, $\alpha = 2\eps$, we get,

\begin{equation}
R^{\, \textit{tor}}_{m,n} = - \frac{\eps}{\prod_{i j} 
\ll i \, b + j \, b^{\, \prime} \rr} 
\prod_{(k,l) \neq(1,1)} \ll 
\frac{1-k}{2} \, b + \frac{1-l}{2} \, b^{-1} \rr + \cO \ll \eps^2 \rr
\end{equation}

\noindent We can call $\rho = \frac{1-k}{2}$ and $\sigma = \frac{1-l}{2}$, then $\rho$ goes from 
$1-m, 2-m, \cdots, m$ and $\sigma$ from $1-n, 2-n, \cdots,n$. 
$(k,l) \neq (1,1) \Leftrightarrow (\rho,\sigma) \neq (0,0)$ and we get,

\begin{multline}
R^{\, \textit{tor}}_{m,n} =  
- \eps \, 
\ll 
\prod_{(i,j) \neq (0,0), (i,j) \neq (m,n)} 
\ll i b + \frac{j}{b} \rr 
\rr^{\, -1} \ll
\prod_{(\rho,\sigma) \neq (0,0)}
\ll \rho b + \frac{\sigma}{b} \rr + \cO \ll \eps^2 \rr \rr
\\
= - \eps \ll m \, b + n \, b^{\, \prime}\rr +\cO \ll \eps^2 \rr 
\end{multline}

\noindent This also implies that,

\begin{equation}
\lim_{\eps \to 0}\frac{R_{\, m, n}\ll 2\eps\rr}{\eps \ll m b + n b^{-1} \rr} = -1
\end{equation}

\noindent for all $b$, for all $(m,n)$.
When $c\in\mathds{Q}$, it can happen that the denominator $r_{m,n}$ vanishes. We will show that the coefficients $R_{m,n}\ll 2\eps \rr$ are always well-defined and express them in closed form. Let's first examine the space $(m,n)$ for which $r_{m,n}\sim\eps^d$.
Let us start with the case $d = 1$. $r_{m,n} = \eps \Rightarrow \exists (i_1,j_1)$ such that,

\begin{equation}
b^2 = -\frac{j_1}{i_1}, 
\quad
1-m\leq i_1 \leq m, 
\quad
1-n\leq j_1 \leq n, 
\quad 
(i_1,j_1) \neq (m,n), 
\end{equation}

\noindent so that,  

\begin{equation}
r_{m,n} = \eps, 
\quad 
(m, n) \in\{p\}\times]\pp,\infty[ \;\bigcup\; ]p,\infty[\times\{\pp\}
\end{equation}

\noindent In the same way, $r_{m,n} = \eps^2 \Rightarrow \exists(i_1,j_1),\;(i_2,j_2)$ such that $b^2 = -\frac{j_1}{i_1} = -\frac{j_2}{i_2}$ with $(i_1,j_1) \neq (i_2,j_2)$, that is, $(i_1,j_1) = (p,\pp)$ and $(i_2,j_2) = (-p,-\pp)$, that is, 

\begin{equation}
r_{m,n} = \eps^2, 
\quad 
(m,n)\in]p, \infty [\times] \pp,2\pp [ \;\bigcup \;] p, 2 p [\times] \pp, \infty[
\end{equation}

\noindent We can generalize to $r_{m,n} = \eps^d$.

\begin{multline}
(m,n)\in\{\frac{d+1}{2}p\}\times]\frac{d+1}{2}\pp,\infty[ \;\bigcup\; ]\frac{d+1}{2}p,\infty[\times\{\frac{d+1}{2}\pp\}\quad d \text{ odd}
\\
(m,n)\in]\frac{d}{2}p,\infty[\times]\frac{d}{2}\pp,(\frac{d}{2}+1)\pp[\;\bigcup\;]\frac{d}{2}p,(\frac{d}{2}+1)p[\times]\frac{d}{2}\pp,\infty[ \quad d \text{ even}
\end{multline}

\begin{figure}[ht]
\centering
\includegraphics[scale=0.8]{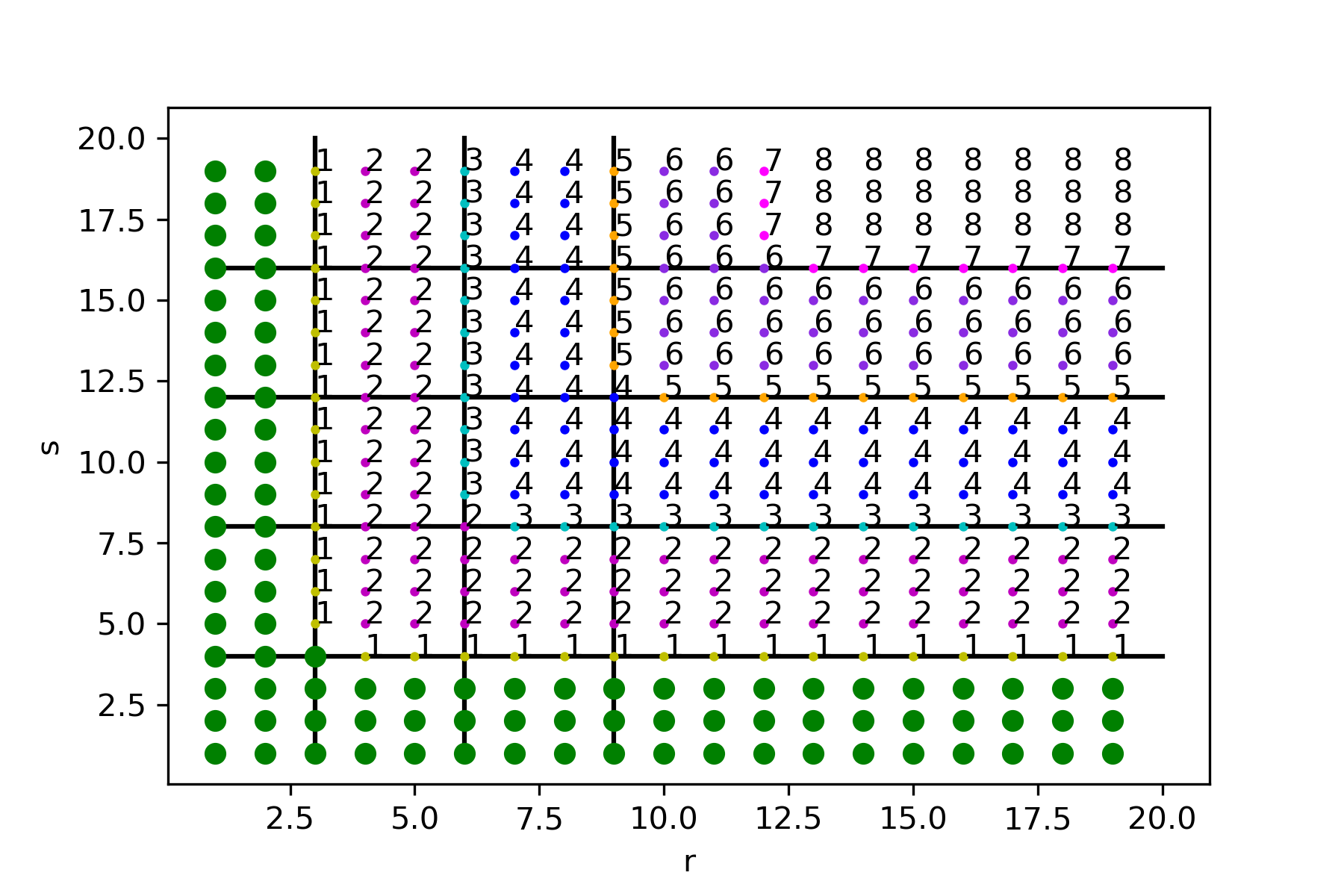}
\caption{Degree $d$ for the Ising model $p = 4$, $\pp = 3$. Green dots are the values 
for which $D_{m, n} \neq 0$}
\label{fig:Dmn}
\end{figure}

\noindent This space is shown on Figure \ref{fig:Dmn} for the Ising model.
Notice that odd $d$ corresponds to the borders of the cells (the fundamental cell being the Kac table). The physical states are not located on the borders, so under the condition that $R^{\, \textit{tor}}_{m,n}\ll2\eps \rr$ vanishes for odd $d$, we can restrict only to even $d$.
For even $d$, the coefficient $R^{\, \textit{tor}}_{m,n}$ is,

\begin{multline}
R^{\, \textit{tor}}_{m,n} \ll 2 \eps \rr = 
\\
- 2 \eps 
\ll m \sqrt{-\frac{\pp}{p}} + \frac{n}{\sqrt{-\frac{\pp}{p}}} +\cO \ll \eps \rr \rr
\times 
\prod_{k = 1}^{\frac{d}{2}}
b^{-1}\ll \pm k p b^2 \pm k\pp - 2 \eps b\rr 
\times
\prod'\ll \rho b + \sigma b^{-1} -2\eps\rr
\\
\times
\frac{
1
}{
2 \prod_{k = 1}^{\frac{d}{2}}
b^{-1}\ll \pm k p b^2 \pm k\pp \rr 
\times
\prod'\ll \rho b + \sigma b^{-1} \rr
}
\end{multline}

\noindent where $\prod'$ is $\prod_{\rho=1-m}^m\prod_{\sigma=1-n}^n$ with $(0,0),(m,n)$ and $(\pm kp,\pm k\pp)$ excluded. Using the regularization for the $b$, 

\begin{equation}
b = \sqrt{-\frac{\pp}{p} \ll 1 + \frac{4\eps}{\sqrt{-p\pp}} \rr}
\end{equation}

\noindent we have, 

\begin{multline}
\pm kp b^2 \pm k\pp = \pm 4 \eps \, k \, \sqrt{-\frac{\pp}{p}} + \cO \ll \eps^2 \rr, 
\\ 
\pm k p b^2 \pm k\pp -2\eps b = 2 \eps \, \sqrt{-\frac{\pp}{p}}\ll \pm 2k - 1\rr + \cO \ll \eps^2 \rr 
\label{terms}
\end{multline}

\noindent which gives,

\begin{equation}
R^{\, \textit{tor}}_{m,n}\ll 2 \eps \rr = - \eps \ll m \, \sqrt{-\frac{\pp}{p}} + \frac{n}{\sqrt{-\frac{\pp}{p}}} \rr 
\times
\frac{1}{2^d}\prod_{k = 1}^{\frac{d}{2}}\frac{4k^2 - 1}{k^2}\; + \cO \ll \eps^2 \rr 
\end{equation}

\noindent for all $(m,n)$ and $d$. For given $(m,n)$, we can write $m = l_m p + m_0$, $n = l_n \pp + n_0$ with $0\leq m_0 < p$ and $0\leq n_0 < \pp$. Then,

\begin{equation}
d = 2 \, \text{min} \ll l_m,l_n \rr \equiv 2 \, l
\label{d}
\end{equation}

\noindent and we can write 

\begin{equation}
R^{\, \textit{tor}}_{m,n} \ll 2 \eps \rr = - \eps \ll m \, \sqrt{-\frac{\pp}{p}} + \frac{n}{\sqrt{-\frac{\pp}{p}}} \rr 
\times
\frac{1}{2^{2l}}\prod_{k = 1}^{l}\frac{4k^2 - 1}{k^2}\; + \cO \ll \eps^2 \rr 
\label{Rmn.02}
\end{equation}

\noindent $l = \text{min} \ll l_m,l_n \rr$.
We need to check that the coefficient also vanishes when $d$ is odd. In that case,

\begin{multline}
R^{\, \textit{tor}}_{m,n} \ll 2 \eps \rr = 
- \eps 
\ll m \sqrt{-\frac{\pp}{p}} + \frac{n}{\sqrt{-\frac{\pp}{p}}} +\cO \ll \eps \rr \rr
\times\ll \frac{d+1}{2}p b^2 + \frac{d+1}{2} \pp -2\eps b\rr \\ \times 
\prod_{k = 1}^{\frac{d-1}{2}}
b^{-1}\ll \pm k p b^2 \pm k\pp - 2 \eps b\rr 
\times
\prod'\ll \rho b + \sigma b^{-1} -2\eps\rr
\\
\times
\frac{
1
}{
\ll \frac{d+1}{2}p b^2 + \frac{d+1}{2}\pp \rr\prod_{k = 1}^{\frac{d-1}{2}}
b^{-1}\ll \pm k p b^2 \pm k\pp \rr 
\times
\prod'\ll \rho b + \sigma b^{-1} \rr
}
\end{multline}

\noindent Using equation (\ref{terms}) we get,

\begin{equation}
R^{\, \textit{tor}}_{m,n}\ll 2\eps \rr = -\eps \ll m \sqrt{-\frac{\pp}{p}} + \frac{n}{\sqrt{-\frac{\pp}{p}}}\rr \frac{d}{2^{d-1}(d+1)}\prod_{k = 1}^{\frac{d-1}{2}}\frac{4k^2-1}{k^2} + \cO \ll \eps^2 \rr 
\end{equation}

\noindent Here $d = 2 \, \text{min}\ll l_m, l_n \rr - 1 \equiv 2 \, l - 1$ which yields,

\begin{equation}
R^{\, \textit{tor}}_{m,n}\ll 2\eps \rr = -\eps \ll m \sqrt{-\frac{\pp}{p}} + \frac{n}{\sqrt{-\frac{\pp}{p}}}\rr \frac{2l-1}{2^{2l-1}\,l}\prod_{k = 1}^{l-1}\frac{4k^2-1}{k^2} + \cO \ll \eps^2 \rr 
\end{equation}

\noindent These terms thus do not contribute in the computation of the character of a physical field.
Now we can figure out expressions for the terms in the recursion.
The non-vanishing terms that involve $\Delta_{int} = \Delta^\eps_{m,n}$ are of the type $\frac{R^{\, \textit{tor}}_{r,s}\ll 2\eps \rr}{\Delta^\eps_{m,n}-\Delta_{r,s}}$ with $\ll m, n \rr \rightarrow \ll r, s \rr^{\pm}_{l}$
Using the regularization of $b$, we have,

\begin{equation}
\Delta^\eps_{m, n} - \Delta_{r, s} = (\pm 2l + 1)\ll m\sqrt{-\frac{\pp}{p}} + \frac{n}{\sqrt{-\frac{\pp}{p}}}\rr \eps + \cO \ll \eps^2 \rr 
\end{equation}

\noindent and using expression (\ref{Rmn}) with $\text{min} \ll l_r,l_s \rr = l -\frac{1}{2} \pm \frac{1}{2}$,  we get, 

\begin{equation}
\lim_{\eps \to 0}\frac{R^{\, \textit{tor}}_{r,s} \ll 2\eps \rr}{\Delta^\eps_{m,n}-\Delta_{r,s}} = -\frac{1}{2^{2(l-\frac{1}{2}\pm \frac{1}{2})}(2l\pm1)}\prod_{k = 1}^{l-\frac{1}{2} \pm \frac{1}{2}}\frac{4k^2-1}{k^2}
\end{equation}

\noindent The terms involving the extra poles are of the form $\frac{R^{\, \textit{tor}}_{r,s}}{\Delta_{m',-n'}-\Delta_{r,s}}$ when $\ll m', -n'\rr \to \ll r, s \rr^\pm_{l'}$. We get,

\begin{equation}
\lim_{\eps \to 0} \frac{\, R^{\, \textit{tor}}_{r,s}}{\Delta_{m',-n'}-\Delta_{r,s}} = -\frac{1}{2^{2l+1}\,l'}\prod_{k = 1}^{l}\frac{4k^2-1}{k^2},\quad l = \text{min}(l_r,l_s).
\end{equation}

\textit{E-mail address:} \texttt{
Nina.Javerzat@lptms.u-psud.fr, Raoul.Santachiara@lptms.u-psud.fr,\newline omar.foda@unimelb.edu.au
}
\end{document}